\documentclass{crc-book}

\usepackage{float}
\usepackage{amsthm,,amssymb, color, epsfig, graphics,bm}
\usepackage{xcolor}
\usepackage{graphicx}
\usepackage[intlimits]{amsmath}
\usepackage{latexsym}

\newtheorem{theorem}{Theorem}
\newtheorem{lemma}{Lemma}
\newtheorem{proposition}{Proposition}

\theoremstyle{definition}

\newtheorem{corollary}{Corollary}

\begin{document}

%
\renewcommand{\evenhead}{Oksana Bihun}
\renewcommand{\oddhead}{Solvable dynamical systems and isospectral matrices}

%
\thispagestyle{empty}

\Name{
Solvable dynamical systems and isospectral matrices defined in terms of the zeros of orthogonal or otherwise special polynomials}

\Author{Oksana Bihun} 

\Address{Department of Mathematics, University of Colorado, Colorado Springs, 1420 Austin Bluffs Pkwy, Colorado Springs 80918, USA}

\begin{abstract}
  \noindent
Several recently discovered properties of multiple families of special polynomials (some orthogonal and some not) that satisfy certain differential, difference or $q$-difference equations are reviewed. A general method of construction of isospectral matrices defined in terms of the zeros of such polynomials is discussed. The method involves reduction of certain partial differential, differential difference and differential $q$-difference equations to systems of ordinary differential equations and their subsequent linearization about the zeros of polynomials in question. Via this process, solvable (in terms of algebraic operations) nonlinear first order systems of ordinary differential equations are constructed.
\end{abstract}


\section{Introduction}


Orthogonal or otherwise special polynomials play an important role in mathematical physics,  for example, in construction of exact solutions of quantum mechanical systems~\cite{NikiforovUvarov88}. They are indispensable in numerical analysis, in part due to the best approximation properties of their linear combinations in certain $L^2$ spaces and the efficiency of the spectral methods for solving differential equations based on these types of approximation~\cite{Phillips03, Funaro92}.
Zeros of special polynomials are significant in many ways, for example, as equilibria of important solvable $N$-body problems~\cite{Calogero2001} or as the nodes in numerical quadrature formulas that yield higher degree of exactness compared to other nodes~\cite{Szego39, Chihara78, MastroianniMilovanovic08}. The zeros of a polynomial $P_N$ from a family $\{P_n\}_{n=0}^\infty$ orthogonal with respect to a positive measure have remarkable properties, in particular, they are distinct and real, remain in the support of the measure, and interlace with the zeros of $P_{N+1}$~\cite{Szego39}. 

The interest in algebraic properties of the zeros of special polynomials dates back to 1885, when Stieltjes established  algebraic relations satisfied by the zeros of classical orthogonal polynomials (Jacobi, Laguerre and Hermite)~\cite{Stieltjes1885-1,Stieltjes1885-2, Stieltjes1885-3}. For example, the zeros $\zeta_n$ of the Hermite polynomial $H_N$ satisfy 
$$
\sum_{j=1, j \neq n}^N\frac{1}{\zeta_j-\zeta_n}=-\zeta_n, \;\; n=1,2,\ldots, N.
$$
The last family of algebraic relations has an electrostatic interpretation: it is  the extremum condition of an energy functional for the system of $N$ unit charges placed on the real line, interacting pairwise according to a repulsive logarithmic potential in the harmonic field~\cite{Stieltjes1885-1,Stieltjes1885-2, Stieltjes1885-3, Szego39, Veselov01}. Other electrostatic models for zeros of polynomials are surveyed in~\cite{MMFMG07}, see also~\cite{Ismail00-1,Ismail00-2}.

Since the groundbreaking work of Stieltjes,  the literature on algebraic relations satisfied by the zeros of orthogonal or otherwise special polynomials is abundant, see for example~\cite{ABCOP1979, Ahmed78, Ahmed79, Calogero77-1, Calogero77-2, Sasaki15, AliciTaseli15, BihunCalogero14-1, BihunCalogero14-2, BihunCalogero15, BihunCalogero16, Bihun17, BihunMourning18}. 

In this chapter we describe a general approach to construction of isospectral  matrices defined in terms of the zeros of special polynomials. The spectrum of these matrices is independent of the zeros and possibly some of the parameters of the polynomials, thus the use of the term ``isospectral'' to describe them. The upshot is that given a polynomial $P_N$ of degree $N$ with the zeros $\bm{\zeta}=(\zeta_1,\ldots,\zeta_N)$, construction of an  $N \times N$ matrix $L\equiv L(\bm{\zeta})$ with the eigenvalues $\lambda_{j}$ independent of the zeros  $\bm{\zeta}$ gives rise to multiple algebraic identities satisfied by these zeros, the identities $\operatorname{Trace} L=\sum_{j=1}^N \lambda_j$ and $\det L= \prod_{j=1}^N \lambda_j$ among them. This approach relies on the differential, difference or $q$-difference equations satisfied by the polynomial $P_N$ and does not use the (possible) orthogonality properties of the family of polynomials that contains $P_N$. Therefore, the results for orthogonal polynomials stated here remain true even if the parameters of these polynomials are outside of the orthogonality range.

The method  goes back to the ideas of Stieltjes and Szeg\"{o}, which exploit the nonlinear relation between time-dependent coefficients $\bm{c}(t)=\left( c_1(t), \ldots, c_N(t) \right)$ and zeros $ \bm{z}(t)=\left( z_1(t), \ldots, z_N(t) \right)$  of a monic polynomial  
\begin{equation}
\psi_N(z,t)=z^N+\sum_{j=1}^N c_j(t) z^{N-j}=\prod_{j=1}^N [z-z_j(t)].
\label{eq1}
\end{equation}
Note the abuse of notation as $z$ is a complex variable in the polynomial $\psi_N(z,t)$, while $\bm{z}(t)$ is a $t$-dependent vector of its zeros.
 Numerous solvable, in terms of algebraic operations, $N$-body problems have been constructed by assuming that $\psi_N(z,t)$ satisfies a linear partial differential equation (PDE)~\cite{Calogero2001}. By recasting the PDE as a linear, hence solvable, system for the coefficients $\bm{c}(t)$ of the polynomial $\psi_N(z,t)$ and, on the other hand, a nonlinear system for the zeros $\bm{z}(t)$ of the same polynomial, one concludes that the nonlinear system is solvable. Indeed,  solutions $\bm{z}(t)$ of the nonlinear system can be recovered via the algebraic operation of finding the zeros  of the $t$-dependent polynomial $\psi_N(z,t)$ with the coefficients $\bm{c}(t)$ that solve the corresponding linear system. If the solution $\bm{c}(t)$ of the linear system is isochronous, i.e. periodic with the period independent of the initial conditions, then the solution $\bm{z}(t)$ of the corresponding nonlinear system is isochronous. This observation is instrumental in construction of many  solvable isochronous systems with remarkable properties~\cite{Calogero08}. Similar constructions of solvable nonlinear dynamical systems have been carried out by comparing the evolution of time-dependent nodes and (generalized or standard) Lagrange basis coefficients of a time-dependent polynomial that satisfies a linear PDE~\cite{Calogero2001}.
 
 Recently, convenient formulas that relate the derivatives of the coefficients $\bm{c}(t)$ with the derivatives of the zeros $\bm{z}(t)$ of the polynomial $\psi_N(z,t)$ have been discovered~\cite{Calogero15, CalogeroBruschi16}. This has expanded the possibilities for construction of solvable nonlinear dynamical systems~\cite{BihunCalogero16-1, BihunCalogero16-2}, for example, see~\cite{BihunCalogero16-2} for a new solvable $N$-body problem that may be interpreted as a hybrid between the Calogero-Moser~\cite{Calogero71, Moser75} and the goldfish~\cite{Calogero01} systems. Using the approach outlined in these recent developments, one may assume that the coefficients $\bm{c}(t)$ of the time-dependent monic polynomial $\psi_N(z,t)$ satisfy a solvable nonlinear dynamical system (as opposed to a linear one, a constraint of the method described above). Then, as before, the corresponding nonlinear system satisfied by the zeros $\bm{z}(t)$ is solvable. Upon iteration of this procedure, one may in fact construct infinite hierarchies of solvable dynamical systems~\cite{BihunCalogero16-3, BihunCalogero17}, a remarkable find given that integrable or solvable dynamical systems are rare  (Poincar\'{e} showed that the integrability of Hamiltonian systems is not a generic property). 

In this chapter we will review construction of several solvable (in terms of algebraic operations) first order systems of nonlinear differential equations of the form 
\begin{equation}
\dot{\bm{z}}(t)=F\big(\bm{z}(t)\big),
\label{zSystemIntro}
\end{equation}
 where $\bm{z}(t)=\big(z_1(t), \ldots, z_N(t) \big)$ and the ``dot'' denotes differentiation with respect to the time-variable $t$. The aim  is to obtain isospectral matrices defined in terms of the zeros of certain special polynomials. To this end, an equation 
 \begin{equation}
 \psi_t=\mathcal{A} \psi
 \label{psiSystemIntro}
 \end{equation} that admits a polynomial solution $\psi_N(z,t)$ of the form~\eqref{eq1}  is constructed, with $\mathcal{A}$ being a linear differential, difference or $q$-difference operator. An additional requirement is that the last equation~\eqref{psiSystemIntro} has a $t$-independent solution $\psi(z,t)=P_N(z)$ with $P_N(z)$ being the $N$-th member of a polynomial family of interest. This ensures that every vector $\bm{\zeta}=\big(\zeta_1,\ldots,\zeta_N\big)$ of the zeros of $P_N(z)$  is an equilibrium of the nonlinear system~\eqref{zSystemIntro} (there are $N!$ of such vectors if the zeros $\zeta_n$ are distinct). The polynomial $\psi_N(z,t)$ solves the equation $\psi_t=\mathcal{A} \psi$ if and only if its coefficients $\bm{c}(t)=\big(c_1(t), \ldots, c_N(t) \big)$ solve a linear system 
 \begin{equation}
 \dot{\bm{c}}(t)=A \bm{c}(t)
 \label{cSystemIntro}
 \end{equation} with the $N\times N$ coefficient matrix $A$, if and only if its zeros $\bm{z}(t)$ solve  system~\eqref{zSystemIntro} with an appropriate right-hand side $F\big(\bm{z}(t)\big)$. A conclusion is that the last system \eqref{zSystemIntro} is solvable since its solutions $\bm{z}(t)$ are the zeros of the polynomial $\psi_N(z,t)$ given by~\eqref{eq1} with the coefficients  $\bm{c}(t)$ that solve system~\eqref{cSystemIntro}.  If the $N\times N$ coefficient matrix of the linear system is upper or lower diagonal, its eigenvalues $\{\lambda_j\}_{j=1}^N$  are easily found. One may then argue that  the  eigenvalues of the coefficient matrix $L=L(\bm{\zeta})$ of the linearization of  system~\eqref{zSystemIntro} about the equilibrium ${\bm{\zeta}}$  are the same and moreover do not depend on some of the parameters of the polynomial $P_N(z)$. The main result is that the matrix $L(\bm{\zeta})$  defined in terms of the zeros  ${\bm{\zeta}}$ of  $P_N(z)$ is isospectral. 

A crucial step in this method is  construction of a linear  operator $\mathcal{A}$ such that equation~\eqref{psiSystemIntro} admits polynomial solutions and possesses a $t$-independent equilibrium solution $P_N(z)$. To accomplish this task, the corresponding differential, difference or $q$-difference equation satisfied by the special polynomial $P_N$ is utilized.

The Askey and the $q$-Askey schemes contain  polynomials orthogonal with respect to a positive measure supported on the real line, which satisfy certain second order linear differential, difference or $q$-difference equations~\cite{KoekoekSwarttouw98}. All these polynomials satisfy certain standard (yet remarkable) properties: three term recurrence relations, Rodrigues-type formulas, formulas that represent their derivatives as linear combinations of polynomials from the same type of families  with shifted parameter(s), explicit formulas for the generating functions and other. Moreover, by taking certain limits with respect to parameters of the polynomials positioned in higher levels of the Askey or the $q$-Askey scheme hierarchy, one obtains polynomials in the lower levels. The Wilson and the Racah polynomials are at the top of the Askey scheme, while the Askey-Wilson and the $q$-Racah  polynomials are at the top of the $q$-Askey scheme, thus our interest in proving algebraic identities for the zeros of these polynomials. All the polynomials in the Askey scheme can be written as special cases of the generalized hypergeometric function, while all the polynomials in the $q$-Askey scheme can be written as particular cases of the generalized basic hypergeometric function. The last two functions play a very important role in mathematics. 

The generalized hypergeometric function can be defined as an infinite series $\sum_{j=0}^\infty A_j z^j$ such that for each $j$, the ratio $\frac{A_{j+1}}{A_j}$ is a rational function of $j$.  The following is a standard notation for the generalized hypergeometric function: 
\begin{eqnarray}
&&{}_{p+1}F_{q}\left( \alpha _{0},\alpha _{1},\ldots,\alpha _{p};\beta
_{1},\ldots,\beta _{q};z\right)\notag\\
&& \equiv _{p+1}F_{q}
\left( \left. 
\begin{array}{c}
\alpha _{0},\alpha _{1},\ldots,\alpha _{p} \\ 
\beta
_{1},\ldots,\beta _{q}
\end{array}
\right\vert z\right)\notag\\
&& =\sum_{j=0}^{\infty }\left[ \frac{\left( \alpha
_{0}\right) _{j}~\left( \alpha _{1}\right) _{j}\cdot \cdot \cdot \left(
\alpha _{p}\right) _{j}~z^{j}}{j!~\left( \beta _{1}\right) _{j}\cdot \cdot
\cdot \left( \beta _{q}\right) _{j}}\right],  
\label{GHypergF}
\end{eqnarray}%
where $(\alpha)_j$ is the Pochhammer symbol  defined  by
\begin{equation*}
\left( \alpha \right) _{0}=1;~~\left( \alpha \right) _{j}=\alpha ~\left(
\alpha +1\right) \cdot \cdot \cdot \left( \alpha +j-1\right) =\frac{\Gamma
\left( \alpha +j\right) }{\Gamma \left( \alpha \right) }~~~\text{for}%
~~~j=1,2,3,\ldots  \label{Poch}
\end{equation*} and $\Gamma$ is the gamma function.
Here,  $\{\alpha_j\}_{j=0}^p$ and $\{\beta_k\}_{k=1}^q$ are complex numbers such that, after appropriate cancellations, the denominators in the series in~\eqref{GHypergF} do not vanish (for example, each $\beta_k$ is not zero and is not a negative integer). Note that if one of the parameters $\alpha_j=-N$ is a negative integer, then series~\eqref{GHypergF} becomes a finite sum and the corresponding generalized hypergeometric function is a polynomial of degree at most $N$ in $z$. In particular, we define the $N$-th generalized basic hypergeometric polynomial  by 
\begin{eqnarray}
&&P_{N}\left( \alpha _{1},\ldots,\alpha _{p};\beta _{1},\ldots,\beta _{q};z\right)
=\sum_{m=0}^{N}\left[ \frac{\left( -N\right) _{m}\left( \alpha _{1}\right)
_{m}\cdot \cdot \cdot \left( \alpha _{p}\right) _{m}~z^{N-m}}{m!~\left(
\beta _{1}\right) _{m}\cdot \cdot \cdot \left( \beta _{q}\right) _{m}}\right]\notag
\\
&=&z^{N}~_{p+1}F_{q}\left( -N,\alpha _{1,}\ldots,\alpha _{p};\beta
_{1},\ldots,\beta _{q};1/z\right). \label{GHypergPol}
\end{eqnarray}
In the case where series~\eqref{GHypergF} is not a polynomial, its radius of convergence $\rho$ can be determined by the ratio test: $\rho=+\infty$ if  $p<q$, $\rho=1$ if $p=q$ and $\rho=0$ if $p>q$, see~\cite{KoekoekLeskySwarttouw10, Erdelyi53}. Of particular interest is the linear differential equation satisfied by the generalized basic hypergeometric function~\eqref{GHypergF}:
\begin{eqnarray}
\left[\mathcal{D}\prod_{j=1}^q (\mathcal{D}+\beta_j-1) -z \prod_{j=0}^p (\mathcal{D}+\alpha_j)\right] u(z)=0,
\label{GHypergFODE}
\end{eqnarray}
where $\mathcal{D}=z~d/dz$, see~\cite{BealsSzmigielski13}. The generalized hypergeometric polynomial $P_N$, on the other hand,  satisfies the following differential equation:
\begin{eqnarray}
\left[\mathcal{D}_N\prod_{j=1}^q (\beta_j-1-\mathcal{D}_N) -\frac{d}{dz} \prod_{j=1}^p (\alpha_j-\mathcal{D}_N)\right] u(z)=0,
\label{GHypergPolODE}
\end{eqnarray}
where 
\begin{equation}
\mathcal{D}_N=z~d/dz-N,
\label{DN}
\end{equation} 
see Section 3 of~\cite{BihunCalogero14-1}  for a proof.

The generalized basic hypergeometric, or $q$-hypergeometric, function with the basis $q \neq 1$ can be defined as an infinite series $\sum_{j=0}^\infty A_j z^j$ such that for each $j$, the ratio $\frac{A_{j+1}}{A_j}$ is a rational function of $q^j$.  A standard notation for the generalized basic hypergeometric function is the following:
\begin{eqnarray}
\label{GBasicHypergF}
&&_{r+1}\phi _{s}\left( \alpha _{0},\alpha _{1,}\ldots,\alpha _{r};\beta
_{1},\ldots,\beta _{s};q;z\right) \\
&& \equiv {}_{r+1} \phi_{s}
\left( \left. 
\begin{array}{c}
\alpha _{0},\alpha _{1},\ldots,\alpha _{r} \\ 
\beta
_{1},\ldots,\beta _{s}
\end{array}
\right\vert q; z\right)\notag\\
&&=\sum_{m=0}^{\infty }\left\{ \frac{\left(
\alpha _{0};q\right) _{m}\left( \alpha _{1};q\right) _{m}\cdot \cdot \cdot
\left( \alpha _{r};q\right) _{m}}{\left( q;q\right) _{m}~\left( \beta
_{1};q\right) _{m}\cdot \cdot \cdot \left( \beta _{s};q\right) _{m}}~\left[
\left( -1\right) ^{m}~q^{m\left( m-1\right) /2}\right] ^{s-r}~z^{m}\right\},
\end{eqnarray}
where  $\{\alpha_j\}_{j=0}^r$ and $\{\beta_k\}_{k=1}^s$ are complex parameters such that, after appropriate cancellations, the denominators in the terms of the above series do not vanish  and 
$$(\gamma;q)_0=1, \;(\gamma;q)_m=(1-\gamma)(1-\gamma q) \cdots(1-\gamma q^{m-1} )\mbox{ for } m=1,2,3,\ldots$$
is the $q$-Pochhammer symbol. 

The generalized basic hypergeometric function is a $q$-analogue of the generalized hypergeometric function in the following sense~\cite{KoekoekLeskySwarttouw10}:
\begin{eqnarray}
  \lim_{q\to 1}   {}_{r+1} \phi_{s}
\left( \left. 
\begin{array}{c}
q^{\alpha _{0}},q^{\alpha _{1}},\ldots,q^{\alpha _{r}} \\ 
q^{\beta_{1}},\ldots,q^{\beta _{s}}
\end{array}
\right\vert q; (q-1)^{s-r}z\right)=    
{}_{r+1} F_{s} 
\left( \left. 
\begin{array}{c}
\alpha_0, \alpha_1, \ldots, \alpha_r\\
\beta_1, \ldots, \beta_s
\end{array}
\right\vert z\right).
\end{eqnarray}

In general, the radius of convergence $\rho$ of the series~\eqref{GBasicHypergF} is  $\rho=\infty$ if $r<s$, $\rho=1$ if $r=s$, and $\rho=0$ if $r>s$, see~\cite{KoekoekLeskySwarttouw10}.
 However, for some choices of the parameters, the series  reduces to a finite sum, producing a polynomial.  Indeed,   if $m>N$,
$(q^{-N};q)_m=(1-q^{-N}) \cdots (1-q^{-N} q^N)\cdots (1-q^{-N} q^{m-1})=0$,
hence $_{r+1}\phi _{s}\left( q^{-N},\alpha _{1,}\ldots,\alpha _{r};\beta
_{1},\ldots,\beta _{s};q;z\right)$ is a {polynomial} of degree at most $N$. Motivated by this observation,  define the $N$-th generalized basic hypergeometric polynomial by~\cite{BihunCalogero15}
\begin{eqnarray}
&& P_{N}\left( \alpha _{1},\ldots,\alpha _{r};\beta _{1},\ldots,\beta
_{s};q;z\right) \notag \\
&&=\sum_{m=0}^{N}\left[ \frac{\left( q^{-N};q\right)
_{m}\left( \alpha _{1};q\right) _{m}\cdot \cdot \cdot \left( \alpha
_{r};q\right) _{m}}{\left( q;q\right) _{m}~\left( \beta _{1};q\right)
_{m}\cdot \cdot \cdot \left( \beta _{s};q\right) _{m}}\left[ \left(
-1\right) ^{m}~q^{\frac{m\left( m-1\right)}{2}}\right] ^{s-r}~z^{m}\right] \notag \\
&&={}_{r+1}\phi _{s}\left( q^{-N},\alpha _{1,}\ldots,\alpha _{r};\beta
_{1},\ldots,\beta _{s};q;z\right).
\label{GBasicHypergPol}
\end{eqnarray}

The generalized basic hypergeometric function~\eqref{GBasicHypergF} and the generalized basic hypergeometric polynomial~\eqref{GBasicHypergPol} satisfy  certain $q$-difference equations. To formulate these equations, let us introduce some $q$-difference operators. 

 Recall that for $q\neq 1$, the $q$-derivetive operator $\mathcal{D}_q$  is defined by~\cite{KoekoekLeskySwarttouw10}
\begin{equation}
\mathcal{D}_q f(z)=\left\{
\begin{array}{ll}
\frac{f(z)-f(qz)}{z-qz}  \mbox{ if } z\neq 0,\\
f'(0) \mbox{ if } z=0.
\end{array}
\right.
\end{equation}
This operator generalizes the differentiation operator $d/dz$: If a function $f(z)$ is differentiable at $z$, then $\lim\limits_{q \to 1} \mathcal{D}_q f(z)=d/dz~f(z).$ 
For the purposes of our study, it is convenient to use the $q$-difference operators $\delta_q$ and $\Delta_\gamma$ defined by
\begin{eqnarray}
&&\delta_q f\left( z\right) =f\left( q z\right), \label{delta_qOperator}\\
&&\Delta _{\gamma } f\left( z\right)=\left( \gamma \delta_q -1\right)f\left( z\right)=\gamma f(q z)-f(z).  \label{DeltaOperator}
\end{eqnarray}
Note that because 
$\Delta _{\gamma } z^{m}=
\left( \gamma q^{m}-1\right)z^{m}$ for every nonnegative integer $m$, 
the action of the operator $\Delta _{\gamma }$ on a polynomial does not raise its degree.

With the above notation, we can state the $q$-difference equation satisfied by the generalized basic hypergeometric function
$_{r+1}\phi _{s}\left( \alpha
_{0},\alpha _{1,}\ldots,\alpha _{r};\beta _{1},\ldots,\beta _{s};z\right) $ (see
Exercise 1.31 on page 27 of \cite{GR1990}): 
\begin{eqnarray}
\Delta _{1}~\left[ \prod\limits_{k=1}^{s}\left( \Delta _{\beta
_{k}/q}\right) \right] u(z) 
-z~\Delta _{q^{-N}}~\left[ \prod\limits_{j=1}^{r}\left( \Delta _{\alpha
_{j}}\right) \right]  u (zq^{s-r}) =0.
\label{qDEGenBasicHyperg}
\end{eqnarray}%
Therefore, the generalized basic hypergeometric polynomial~\eqref{GBasicHypergPol} satisfies the following $q$-difference equation~\cite{BihunCalogero15}:
\begin{equation}
\left[\Delta_1 \prod_{k=1}^s (\Delta_{\beta_k/q} )-z~\Delta_{q^{-N}} \prod_{j=1}^r (\Delta_{\alpha_j}) (\delta_q)^{s-r} \right] u(z)=0.
\label{qDifferenceODEGenBasic}
\end{equation}

\section{Zeros of generalized hypergeometric polynomial with two parameters and zeros of Jacobi polynomials}
\label{PedagogicalExample}


In this section we illustrate the general method of construction of an isospectral matrix defined in terms of the zeros of a special polynomial $P_N(z)$ for the simple example where
\begin{equation}
P_N(z) \equiv P_{N}\left( \alpha _{1};\beta _{1};z\right) =\sum_{m=0}^{N}\left[ \frac{%
\left( -N\right) _{m}\left( \alpha _{1}\right) _{m}~z^{N-m}}{m!~\left( \beta
_{1}\right) _{m}}\right]  \label{PNp=q=1}
\end{equation}
is the generalized basic hypergeometric polynomial~\eqref{GHypergPol} with two parameters $\alpha_1$ and $\beta_1$, see~\cite{BihunCalogero14-1}.
By taking $p=q=1$ in~\eqref{GHypergPolODE}, we conclude that  this polynomial  satisfies the differential equation
\begin{eqnarray}
\mathcal{A} u(z)=0,
\label{ExPolODE}
\end{eqnarray}
where $\mathcal{A}$ is the linear differential operator
\begin{equation}
\mathcal{A}= \sum_{j=1}^2 b_j (\mathcal{D}_N)^j-\frac{d}{dz} \sum_{j=0}^1 a_j (\mathcal{D}_N)^j 
\label{Aoperator}
\end{equation}
with $\mathcal{D}_N=z\frac{d}{dz}-N$ or $\mathcal{D}_N=z\frac{\partial}{\partial z}-N$ depending on the context, see~\eqref{DN}, and
\begin{equation}
a_{0}=\alpha _{1},~a_{1}=-1,~b_{1}=\beta _{1}-1 \mbox{ and } ~b_{2}=-1.
\label{a0a1b1b2}
\end{equation}

To construct an $N\times N$ isospectral matrix defined in terms of the zeros $\bm{\zeta}=(\zeta_1, \ldots, \zeta_N)$ of the polynomial $P_N(z)$,  consider the PDE
\begin{equation}
\frac{\partial }{\partial t} \psi(z,t) =- \mathcal{A} \psi(z,t),
\label{ExPDE}
\end{equation}
which has a time-independent solution $\psi(z,t)=P_N(z)$. Equation~\eqref{ExPDE} admits time-dependent monic polynomial solutions of the form~\eqref{eq1}. Indeed,  upon substitution of the anzatz $\psi_N(z,t)=z^N+\sum_{j=1}^N c_j(t) z^{N-j}$ into~\eqref{ExPDE}, we obtain the following linear system of differential equations for the coefficients $\bm{c}(t)=\left(c_1(t), \ldots, c_N(t) \right)$:
\begin{eqnarray}
\dot{c}_{m}=m
\left( \beta _{1}-1+m\right) c_{m}+\left( N+1-m\right) \left( \alpha
_{1}-1+m\right) c_{m-1},
\label{ExSystc}
\end{eqnarray}
where $m=1,\ldots, N$, $c_{0} =1$ and  $c_j=0$  for all integer $j$ outside of the interval $[0,N]$. The last system can be recast in the form
\begin{equation}
\dot{\bm{c}}(t)=A \bm{c}(t) +\bm{h},
\label{ExSystcMatrixForm}
\end{equation}
where $A$ is a lower diagonal $N \times N$ matrix with the eigenvalues 
\begin{equation}
\lambda_m=m(\beta_1-1+m)
\label{Exlambdam}
\end{equation}
that do not depend on the parameter $\alpha_1$ of the polynomial $P_N(z)$ and $\bm{h}$ in the $N$-vector $\bm{h}=(N\alpha_1,0,\ldots,0)^T$. Assuming that these eigenvalues are all different among themselves,   a general solution of  linear system~\eqref{ExSystcMatrixForm} can be written as
\begin{eqnarray}
\bm{c}(t)=\sum_{j=1}^N \eta_m e^{\lambda_m t} \bm{v}_m+\bm{c}_p,
\label{Excsoln}
\end{eqnarray}
where $\eta_m$ are $N$ arbitrary constants, each  $\bm{v}_m$ is an eigenvector of the matrix $A$ that corresponds to the eigenvalue $\lambda_m$ and $\bm{c}_p=-A^{-1} h$ is a particular solution of~\eqref{ExSystcMatrixForm} (note a misprint in (34) of~\cite{BihunCalogero14-1}).

Let us now substitute the representation $\psi_N(z,t)=\prod_{m=1}^N[z-z_m(t)]$ of the time-dependent polynomial~\eqref{eq1} in terms of its zeros $\bm{z}(t)=\left(z_1(t), \ldots, z_N(t)\right)$ into PDE~\eqref{ExPDE}. In terms of computations, this last substitution is somewhat less trivial, given that our aim is to obtain a first order nonlinear system of ODEs for the zeros $\bm{z}(t)$ of the form~\eqref{zSystemIntro}. 
Before we discuss a convenient method for construction of system~\eqref{zSystemIntro} satisfied by the zeros $\bm{z}(t)$ of the polynomial $\psi_N(z,t)$ satisfying PDE~\eqref{ExPDE}, let us outline a general strategy for construction of an $N \times N$ isospectral matrix $L(\bm{\zeta})$ defined in terms of the zeros $\bm{\zeta}$ of the polynomial $P_N(z)$. 

Because $P_N(z)$ is a time-independent polynomial solution of  PDE~\eqref{ExPDE}, the vector $\bm{\zeta}$ of its zeros is an equilibrium of system~\eqref{zSystemIntro}, that is, $F(\bm{\zeta})=0$. It is therefore natural to linearize this last system about its equilibrium $\bm{\zeta}$ to obtain the system
\begin{equation}
\dot{x}(t)=L x(t),
\label{ExSystxLinearized}
\end{equation}
where the $N\times N$ matrix $L\equiv L(\bm{\zeta})=D F(\bm{\zeta})$ is the Jacobian matrix of the function $F$ in the right-hand side of~\eqref{zSystemIntro}, evaluated at $\bm{\zeta}$. The next step is to compare the eigenvalues of the matrix $L(\bm{\zeta})$ with the eigenvalues $\{\lambda_m\}_{m=1}^N$ of the matrix of coefficients $A$ of system~\eqref{ExSystcMatrixForm} to conclude that these eigenvalues must be equal. A corollary of the last statement is that the $N\times N$ matrix $L(\bm{\zeta})$ defined in terms of the zeros $\bm{\zeta}$ of the polynomial $P_N(z)$ has the $N$ eigenvalues $\lambda_m=m(\beta_1-1+m)$, see~\eqref{Exlambdam}, which depend neither on the zeros $\bm{\zeta}$ nor on the parameter $\alpha_1$ of the polynomial $P_N(z)$ and, moreover, are rational if the parameter $\beta_1$ is rational, a diophantine property.

To construct system~\eqref{zSystemIntro}, we use the following notation, see~\cite{Calogero08}. For a monic polynomial 
\begin{equation}
\phi(z)=\prod_{n=1}^N(z-z_n)
\label{psi}
\end{equation}
 with the vector of zeros $(z_1, \ldots, z_N)$ and a linear differential operator $\mathcal{B}$  with polynomial coefficients, acting in the variable $z$, we write 
\begin{equation}
\mathcal{B}\phi \left( z\right) \longleftrightarrow f_{n}\left( z_1,\ldots, z_N \right)
\end{equation}
to denote the identity
\begin{equation}
\mathcal{B}\phi \left( z\right) =\phi(z) \sum_{n=1}^{N}\left[ \left( z-z_{n}\right)
^{-1}~f_{n}\left(z_1,\ldots, z_N  \right) \right],
\end{equation}
note a misprint in (48b) of~\cite{BihunCalogero14-1}.
For example, by logarithmic differentiation of~\eqref{psi}, one obtains
\begin{equation}
\left( \frac{d}{dz}\right) \phi \left( z\right) \longleftrightarrow 1,
\label{A.4}
\end{equation}
and, from the last identity, by using $z/\left( z-z_{n}\right)
=1+z_{n}/\left( z-z_{n}\right) $, one obtains
\begin{equation}
\mathcal{D}_N \phi \left( z\right) \longleftrightarrow
z_{n},  \label{A.6a}
\end{equation}
where $\mathcal{D}_N$ is given by~\eqref{DN},   see (A.4) and (A.6a) in~\cite{Calogero08}. 

Let us find $f_n^{\mathcal{A}}(z_1,\ldots, z_N)$  such that 
\begin{equation}
\mathcal{A}\phi \left( z\right) \longleftrightarrow f_{n}^{\mathcal{A}}\left( z_1,\ldots, z_N \right).
\label{Anotation}
\end{equation}
A computation of $f_n^{\mathcal{A}}(z_1,\ldots, z_N)$ will accomplish the goal of construction of system~\eqref{zSystemIntro}. Indeed, for the polynomial ~$\psi_N(z,t)=\prod_{n=1}^N[z-z_n(t)]$,
$$-\mathcal{A}\psi_N=-\psi_N  \sum_{n=1}^N [z-z_n(t)]^{-1} f_{n}^{\mathcal{A}}\left( z_1,\ldots, z_N \right)$$
and
 $$\partial \psi_N /\partial t=-\psi_N \sum_{n=1}^N [z-z_n(t)]^{-1}\dot{z_n},$$ 
hence~$\psi_N(z,t)$
solves PDE~\eqref{ExPDE}
 if and only if its zeros $\bm{z}(t)=(z_1(t), \ldots, z_N(t))$ satisfy  the differential equations
\begin{equation}
\dot{z}_n(t)=f_{n}^{\mathcal{A}}\left( z_1(t),\ldots, z_N(t) \right), 
\label{z_nSyst1}
\end{equation}
where $n=1,2,\ldots,N$. System~\eqref{z_nSyst1} is an explicit form of system~\eqref{zSystemIntro}.
 
Motivated by formula~\eqref{Aoperator} for the differential operator $
\mathcal{A}$, we introduce the expressions $f_n^{(j)}(z_1, \ldots, z_N)$ and $g_n^{(j)}(z_1, \ldots, z_N)$ to denote
\begin{equation}
(\mathcal{D}_N)^j~\phi \left( z\right) \longleftrightarrow f_{n}^{(j)}\left( z_1,\ldots, z_N \right)
\label{fjnotation}
\end{equation}
and
\begin{equation}
\frac{d}{dz}(\mathcal{D}_N)^j~\phi \left( z\right) \longleftrightarrow g_{n}^{(j)}\left( z_1,\ldots, z_N \right)
\label{gjnotation}
\end{equation}
so that
\begin{equation}
f_n^{\mathcal{A}}(z_1,\ldots, z_N)=b_1 f_n^{1}(z_1,\ldots, z_N)+b_2 f_n^{2}(z_1,\ldots, z_N)+
a_0 g_n^{0}(z_1,\ldots, z_N)+a_1 g_n^{1}(z_1,\ldots, z_N).
\label{fnA=fnjgnj}
\end{equation}
By~\eqref{A.6a} and~\eqref{A.4}, 
\begin{equation}
f_n^{(1)}(z_1, \ldots, z_N)=z_n \mbox{ and } g_n^{(0)}(z_1, \ldots, z_N)=1.
\label{fn1gn0}
\end{equation}

To compute $f_n^{(2)}(z_1, \ldots, z_N)$, let us apply the differential operator $(\mathcal{D}_N)^2$ to the polynomial $\phi(z)$ defined by~\eqref{psi}:
\begin{eqnarray}
&&(\mathcal{D}_N)^2\phi =\mathcal{D}_N \left[ \phi \sum_{n=1}^N (z-z_n)^{-1} z_n \right]= (\mathcal{D}_N \phi) \sum_{n=1}^N\frac{z_n}{z-z_n}+\phi \sum_{n=1}^N z \frac{d}{dz}\left( \frac{z_n}{z-z_n}\right)\notag\\
&&=\phi \left[ \sum_{n,m=1}^N \frac{z_n z_m}{(z-z_n)(z-z_m)}-\sum_{n=1}^N \frac{(z-z_n) z_n}{(z-z_n)^2} -\sum_{n=1}^N \frac{(z_n)^2}{(z-z_n)^2}  \right] \notag\\
&&=\phi\left[\sum_{n,m=1, n\neq m}^N \frac{z_n z_m}{(z-z_n)(z-z_m)}  -\sum_{n=1}^N \frac{ z_n}{z-z_n}\right]\notag
\end{eqnarray}
\begin{eqnarray}
&&=\phi\left\{\sum_{n,m=1, n\neq m}^N \frac{z_n z_m}{z_n-z_m} \left[\frac{1}{z-z_n}-\frac{1}{z-z_m} \right]  -\sum_{n=1}^N \frac{ z_n}{z-z_n}\right\}\notag\\
&&=\phi \sum_{n=1}^N (z-z_n)^{-1} \left[ \sum_{m=1, m\neq n}^N \frac{2 z_n z_m}{z_n-z_m} -z_n \right].
\end{eqnarray}
Therefore,
\begin{equation}
f_n^{(2)}(z_1, \ldots, z_N)=\sum_{m=1, m\neq n}^N \frac{2 z_n z_m}{z_n-z_m} -z_n.
\label{fn2}
\end{equation}

To compute $g_n^{(2)}(z_1, \ldots, z_N)$, consider
\begin{eqnarray}
&&\frac{d}{dz}( \mathcal{D}_N \phi) =\frac{d}{dz}\left[ \phi \sum_{n=1}^N (z-z_n)^{-1} z_n
\right]=\left(\frac{d \phi}{dz} \right) \sum_{n=1}^N \frac{ z_n}{z-z_n}-\phi \sum_{n=1}^N\frac{z_n}{(z-z_n)^2}\notag\\
&&=\phi \left[ \sum_{n,m=1, n \neq m}^N \frac{z_n}{(z-z_n)(z-z_m)}\right]\notag\\
&&=\phi \left\{ \sum_{n,m=1, n \neq m}^N \frac{z_n}{z_n-z_m} \left[ \frac{1}{z-z_n}-\frac{1}{z-z_m} \right]\right\} \notag\\
&&=\phi \sum_{n=1}^N (z-z_n)^{-1} \left[\sum_{m=1, m \neq n}^N \frac{z_n+z_m}{z_n-z_m}\right].
\end{eqnarray}
Therefore,
\begin{equation}
g_n^{(1)}(z_1, \ldots, z_N)=\sum_{m=1, m \neq n}^N \frac{z_n+z_m}{z_n-z_m}.
\label{gn1}
\end{equation}
Using~\eqref{fnA=fnjgnj}, \eqref{a0a1b1b2}, \eqref{fn1gn0}, \eqref{fn2} and \eqref{gn1}, we find
\begin{eqnarray}
&&f_n^{\mathcal{A}}(z_1,\ldots, z_N)=-\alpha_1+\beta_1 z_n+\sum_{m=1, m\neq n}^N \frac{z_n+z_m-2 z_n z_m}{z_n-z_m}\notag\\
&&=N-1-\alpha_1+\beta_1 z_n +2(1-z_n) \sum_{m=1, m\neq n}^N\frac{z_m}{z_n-z_m}.
\label{fnAExplicit}
\end{eqnarray}

We proved the following Lemma.

\begin{lemma} The time-dependent monic polynomial~\eqref{eq1} satisfies PDE~\eqref{ExPDE} if and only if its coefficients $\bm{c}(t)$ satisfy the linear system~\eqref{ExSystc} if and only if its zeros $\bm{z}(t)$ satisfy the nonlinear system
\begin{equation}
\dot{z}_n(t)=N-1-\alpha_1+\beta_1 z_n(t) +2[1-z_n(t)] \sum_{m=1, m\neq n}^N\frac{z_m(t)}{z_n(t)-z_m(t)}.
\label{z_nSyst1Explicit}
\end{equation}
In particular, system~\eqref{z_nSyst1Explicit} is solvable in terms of algebraic operations.
\end{lemma}

Because $P_N(z)$ given by~\eqref{PNp=q=1} is a time-independent solution of PDE~\eqref{ExPDE}, the vector of its zeros $\bm{\zeta}=(\zeta_1, \ldots, \zeta_N)$ is an equilibrium of system~\eqref{z_nSyst1Explicit}. Therefore, the following corollary holds.

\begin{corollary}
\label{AlgIdentitiesP11}
\textit{If the (possibly complex) zeros $\zeta_1, \ldots, \zeta_N$ of the generalized basic hypergeometric polynomial $P_N(z)=P_N(\alpha_1, \beta_1;z)$ defined by~\eqref{PNp=q=1} are all different among themselves, they satisfy the following family of algebraic identities~\cite{BihunCalogero14-1}:
\begin{equation}
 2(\zeta_n-1) \sum_{m=1, m\neq n}^N\frac{\zeta_m}{\zeta_n-\zeta_m}=N-1-\alpha_1+\beta_1 \zeta_n,
\end{equation}
where $n=1,2,\ldots,N$.}
\end{corollary} 

Let us now linearize system~\eqref{z_nSyst1Explicit} about its equilibrium~$\bm{\zeta}=(\zeta_1, \ldots, \zeta_N)$ to obtain linear system~\eqref{ExSystxLinearized}. The $N\times N$ matrix of coefficients $L \equiv L(\bm{\zeta})$ of system~\eqref{ExSystxLinearized} is given componentwise by 
\begin{equation}L_{nm}=\frac{\partial}{ \partial z_m} f_n^{\mathcal{A}}(z_1, \ldots, z_N)\Big|_{z_j=\zeta_j, \,\, j=1,\ldots,N,}
\label{LnmfnA}
\end{equation}
where $ f_n^{\mathcal{A}}(z_1, \ldots, z_N)$ are given by~\eqref{fnAExplicit}.  Its eigenvalues coincide with the eigenvalues~\eqref{Exlambdam} of the matrix of coefficients $A$ of system~\eqref{ExSystc},~\eqref{ExSystcMatrixForm} (see the proof of the next theorem), therefore, the following Theorem holds.

\begin{theorem}
\label{Thm:IsospMatrixP11} Let $\bm{\zeta}=(\zeta_1, \ldots, \zeta_N)$ be a vector of the zeros of the generalized basic hypergeometric polynomial $P_N(z)=P_N(\alpha_1, \beta_1;z)$ defined by~\eqref{PNp=q=1}. If $\zeta_n$ are all different among themselves, then the $N\times N$ matrix $L\equiv L(\bm{\zeta})$ defined componentwise by~\cite{BihunCalogero14-1}
\begin{eqnarray}
L_{nm}&\equiv &  L_{nm}(\alpha_1, \beta_1; N;\bm{\zeta})=\delta _{nm}~\left\{ \beta
_{1}+2~\sum_{\ell =1;\ell \neq n}^{N}\left[ \frac{\zeta _{\ell }~\left(
\zeta _{\ell }-1\right) }{\left( \zeta _{n}-\zeta _{\ell }\right) ^{2}}%
\right] \right\}  \notag \\
&&-2~\left( 1-\delta _{nm}\right) ~\frac{\zeta _{n}~\left( \zeta
_{n}-1\right) }{\left( \zeta _{n}-\zeta _{m}\right) ^{2}}  \label{12d}
\end{eqnarray}
has the eigenvalues $\lambda_m\equiv \lambda_m(\beta_1;N)=m(\beta_1-1-m)$, where $m=1,2,\ldots, N$.
\end{theorem}
Note that the eigenvalues $\lambda_m$ of the last matrix $L$ do not depend on the parameter $\alpha_1$ and are moreover rational if $\beta_1$ is rational, a diophantine property.
\begin{proof}
Formulas~\eqref{12d} follow from~\eqref{LnmfnA}. Let $\{\tilde{\lambda}_m\}_{m=1}^N$ be the eigenvalues of the matrix $L$. In case these eigenvalues are all different among themselves, a general solution of system~\eqref{ExSystxLinearized} is given by  \begin{eqnarray}
\bm{x}(t)=\sum_{j=1}^N \tilde{\eta}_m e^{\tilde{\lambda}_m t} \tilde{\bm{v}}_m,
\end{eqnarray}
where $\tilde{\eta}_m$ are $N$ arbitrary constants and each  $\tilde{\bm{v}}_m$ is an eigenvector of the matrix $L$ that corresponds to the eigenvalue $\tilde{\lambda}_m$. But the behavior of the solution of the linearization~\eqref{ExSystxLinearized} of system~\eqref{z_nSyst1Explicit} in the vicinity of its equilibrium $\bm{\zeta}$  cannot differ from its general behavior. The solutions $\bm{z}(t)$ of system~\eqref{z_nSyst1Explicit} are the zeros of the monic time-dependent polynomial~\eqref{eq1} such that the behavior of its coefficients $\bm{c}(t)$ is characterized by the $N$ exponential functions $e^{\lambda_m t}$, see~\eqref{Excsoln}. Therefore, the set of eigenvalues $\tilde{\lambda}_m$ of the matrix $L(\bm{\zeta})$ coincides with the set of eigenvalues $\lambda_m$, see \eqref{Exlambdam}, of the matrix of coefficients $A$ of system~\eqref{ExSystc},~\eqref{ExSystcMatrixForm}.\end{proof}

The Jacobi polynomials $P_N^{(\alpha, \beta)}(x)$ can be expressed in terms of the generalized hypergeometric polynomial $P_N(\alpha_1; \beta_1; z)$ defined by~\eqref{PNp=q=1}, see (1.8.1) in~\cite{KoekoekSwarttouw98}:
\begin{equation}
P_N^{(\alpha, \beta)}(x)=\frac{(\alpha+1)_N}{N!} \left( \frac{1-x}{2} \right)^N P_N\left(N+\alpha+\beta+1; \alpha+1; \frac{2}{1-x} \right).
\end{equation}
Therefore, the results of Corollary~\ref{AlgIdentitiesP11} and Theorem~\ref{Thm:IsospMatrixP11} can be recast in terms of the zeros $\bm{x}=(x_1,\ldots, x_N)$ of the Jacobi polynomial $P_N^{(\alpha, \beta)}(x)$.  In this setting, Corollary~\ref{AlgIdentitiesP11}  
reduces to eqs. (5.2a) respectively
(5.2b) of \cite{ABCOP1979}, while Theorem~\ref{Thm:IsospMatrixP11} produces results discovered more recently~\cite{BihunCalogero14-1}, which we state below.
\begin{theorem} If $\bm{x}=(x_1,\ldots, x_N)$ is a vector of distinct zeros of the Jacobi polynomial $P_N^{(\alpha, \beta)}(x)$, then 
the $N\times N$ matrix $L\left( 
\bm{x}\right) $ defined componentwise by~\cite{BihunCalogero14-1}
\begin{eqnarray}
&&L_{nm}\equiv L_{nm}\left(\alpha, \beta;N; \bm{x}\right) =\delta _{nm}~\left\{ \alpha
+1+\sum_{\ell =1,~\ell \neq n}^{N}\left[ \frac{\left( 1+x_{\ell }\right)
~\left( 1-x_{n}\right) ^{2}}{\left( x_{n}-x_{\ell }\right) ^{2}}\right]
\right\}   \notag \\
&&-\left( 1-\delta _{nm}\right) ~\left[ \frac{\left( 1+x_{n}\right) ~\left(
1-x_{m}\right) ^{2}}{\left( x_{n}-x_{m}\right) ^{2}}\right]
\end{eqnarray}%
has the $N$ eigenvalues%
\begin{equation}
\lambda _{m} \equiv \lambda_m(\alpha;N)=m~\left( m+\alpha \right), \;\;\;m=1,\ldots,N.
\end{equation}
\end{theorem}

Note the isospectral property of  the matrix $
{L}\left(\alpha, \beta;N; \bm{x}\right)$: its elements depend on the zeros $\bm{x} \equiv \bm{x} \left( \alpha ,\beta \right) $ and, via these zeros, on the two
parameters $\alpha $ and $\beta $, while its eigenvalues depend only on the
parameter $\alpha$. Moreover, if $\alpha$ is rational, then the eigenvalues $\lambda_m$ are rational, a diophantine property.

Let us also note that in the last Theorem, the parameters $\alpha, \beta$ need not be in the real range $(-1,+\infty)$ that ensures the orthogonality of the Jacobi polynomial family. However, it is assumed that the complex parameters $\alpha, \beta$ are such that the (possibly complex) zeros $x_n$ of $P_N^{(\alpha, \beta)}(x)$ are all different among themselves. This is true, for example, for the quasiorthogonal case where $\alpha>-1$ and $-2<\beta<-1$, see~\cite{DriverJordaan16} and references therein.

\section{Zeros of generalized hypergeometric polynomials}

Using an approach similar to that described in  Section~\ref{PedagogicalExample}, it is possible to prove algebraic identities and to construct  isospectral matrices defined in terms of the zeros of the generalized hypergeometric polynomial~\eqref{GHypergPol}. Before stating these results, let us introduce a few new definitions.

Given  $p$ complex parameters $\alpha_1, \ldots, \alpha_p$ and $q$ complex parameters $\beta_1, \ldots, \beta_q$ of the generalized hypergeometric polynomial $P_{N}\left( \alpha _{1},\ldots,\alpha
_{p};\beta _{1},\ldots,\beta _{q};z\right) $, let the complex numbers $a_0, \ldots, a_{p}$ and $b_1, \ldots, b_{q+1}$ be defined as the unique set of coefficients that makes the following polynomial equations true:
\begin{equation}
\prod\limits_{j=1}^{p}\left( \alpha _{j}-x\right)
=\sum_{j=0}^{p}a_{j}~x^{j},
\label{aj}
\end{equation}
\begin{equation}
x~\prod\limits_{k=1}^{q}\left( \beta _{k}-1-x\right)
=\sum_{k=1}^{q+1}\left( b_{k}~x^{k}\right).
\label{bk}
\end{equation}
Then
\begin{eqnarray}
&&a_{0}=\prod\limits_{j=1}^{p}\left( \alpha _{j}\right), \;\;\;
a_{1}=-\sum_{k=1}^{p}\left[ \prod\limits_{j=1;~j\neq k}^{p}\left( \alpha
_{j}\right) \right], \notag\\
&&a_{2}=\frac{1}{2}\sum_{\ell ,k=1;\ell \neq k}^{p}\left[ \prod%
\limits_{j=1;~j\neq \ell ,k}^{p}\left( \alpha _{j}\right) \right],
\ldots, \;\;\;
a_{p}=\left( -1\right) ^{p}
\end{eqnarray}
and
\begin{eqnarray}
&&b_{1}=\prod\limits_{k=1}^{q}\left( \beta _{k}-1~\right), \;\;\;
b_{2}=-\sum_{j=1}^{q}\left[ \prod\limits_{k=1,~k\neq j}^{q}\left( \beta
_{k}-1\right) \right],\notag\\
&&b_{3}=\frac{1}{2}\sum_{\ell ,j=1;\ell \neq j}^{q}\left[ \prod%
\limits_{k=1,~k\neq \ell ,j}^{q}\left( \beta _{k}-1\right) \right],
\ldots,
b_{q+1}=\left( -1\right) ^{q}.
\end{eqnarray}

Given a vector $\bm{\zeta}=(\zeta_1, \ldots, \zeta_N)$ in $\mathbb{C}^N$ and an integer $n$ such that $1\leq n \leq N$, define the functions $f_n^{(j)}(\bm{\zeta})$ and $g_n^{(j)}(\bm{\zeta})$ as follows:
\begin{eqnarray}
&&f_{n}^{\left( j+1\right) }\left( \bm{\zeta }\right) =-f_{n}^{\left(
j\right) }\left( \bm{\zeta }\right) +\sum_{\ell =1;~\ell \neq n}^{N}%
\left[ \frac{\zeta _{n}~f_{\ell }^{\left( j\right) }\left(  \bm{\zeta }%
\right) +\zeta _{\ell }~f_{n}^{\left( j\right) }\left(  \bm{\zeta }%
\right) }{\zeta _{n}-\zeta _{\ell }}\right],\notag\\
&&f_{n}^{\left( 1\right) }\left( \bm{\zeta }\right) =\zeta _{n},
\label{fnj}
\end{eqnarray}%
and 
\begin{eqnarray}
&&g_{n}^{\left( j\right) }\left( \bm{\zeta }\right) =\sum_{\ell
=1;~\ell \neq n}^{N}\left[ \frac{f_{n}^{\left( j\right) }\left(  \bm{%
\zeta }\right) +f_{\ell }^{\left( j\right) }\left(  \bm{\zeta }\right) 
}{\zeta _{n}-\zeta _{\ell }}\right],\notag\\
&&g_{n}^{\left( 0\right) }\left( \bm{\zeta }\right) =1,
\label{gnj}
\end{eqnarray}
where $j=1,2,\ldots$.
This definition is consistent with the notation~\eqref{fjnotation},~\eqref{gjnotation} of the previous section.

Using the notation
\begin{equation}
\sigma _{n}^{\left( r,\rho \right) }\left( \bm{\zeta}\right) =\sum_{\ell
=1;\ell \neq n}^{N}\frac{(\zeta_{\ell })^{r}}{\left( \zeta_{n}-\zeta_{\ell }\right)
^{\rho }},  \label{sigmanpq}
\end{equation}
one may express the above functions, for small values of $j$,  as follows:
\begin{eqnarray*}
&&f_{n}^{\left( 2\right) }\left(  \bm{\zeta}\right)=\zeta_{n}~\left[ -1+2~\sigma _{n}^{\left( 1,1\right) }\left(  \bm{\zeta}%
\right) \right] ~,  \label{fn2}
\end{eqnarray*}%
\begin{eqnarray*}
&&f_{n}^{\left( 3\right) }\left(  \bm{\zeta}\right) =\zeta_{n}\left\{ 1-6~\sigma _{n}^{\left( 1,1\right) }\left(  \bm{\zeta}%
\right) -3~\sigma _{n}^{\left( 2,2\right) }\left(  \bm{\zeta}\right) +3~%
\left[ \sigma _{n}^{\left( 1,1\right) }\left(  \bm{\zeta}\right) \right]
^{2}\right\} ~,  \label{fn3}
\end{eqnarray*}%
\begin{eqnarray*}
&&f_{n}^{\left( 4\right) }\left(  \bm{\zeta}\right) =\zeta_{n}~\left\{ -1+14~\sigma _{n}^{(1,1)}( \bm{\zeta})+18~\sigma
_{n}^{(2,2)}( \bm{\zeta})+8~\sigma _{n}^{(3,3)}( \bm{\zeta})\right. 
\notag \\
&&\left. -18~\left[ \sigma _{n}^{(1,1)}( \bm{\zeta})\right] ^{2}-12~\sigma
_{n}^{(1,1)}( \bm{\zeta})~\sigma _{n}^{(2,2)}( \bm{\zeta})+4~\left[
\sigma _{n}^{(1,1)}( \bm{\zeta})\right] ^{3}\right\} ,
\end{eqnarray*}%
and
\begin{eqnarray*}
&&g_{n}^{\left( 1\right) }\left( \bm{\zeta}\right) =N-1+2~\sigma _{n}^{\left( 1,1\right) }\left( \bm{\zeta}\right) ~,
\label{gn1}
\end{eqnarray*}
\begin{eqnarray*}
&&g_{n}^{\left( 2\right) }\left( \bm{\zeta}\right) =1-N+2~\left( N-3\right) ~\sigma _{n}^{\left( 1,1\right) }\left( 
\bm{\zeta}\right) -3~\sigma _{n}^{\left( 2,2\right) }\left( \bm{\zeta}%
\right) +3~\left[ \sigma _{n}^{\left( 1,1\right) }\left( \bm{\zeta}%
\right) \right] ^{2}~,  \label{gn2}
\end{eqnarray*}
\begin{eqnarray*}
&&g_{n}^{\left( 3\right) }\left( \bm{\zeta}\right) =N-1-2~\left( 3~N-7\right) \sigma _{n}^{(1,1)}(\bm{\zeta})-\frac{9}{4}%
~\left( N-7\right) ~\sigma _{n}^{(2,2)}(\bm{\zeta})+6~\sigma _{n}^{(3,3)}(%
\bm{\zeta})  \notag \\
&&+\frac{9}{4}~\left( N-7\right) ~~\left[ \sigma _{n}^{(1,1)}(\bm{\zeta})%
\right] ^{2}-9~\sigma _{n}^{(1,1)}(\bm{\zeta})~\sigma _{n}^{(2,2)}(%
\bm{\zeta})+3~\left[ \sigma _{n}^{(1,1)}(\bm{\zeta})\right] ^{3}.
\end{eqnarray*}%

\begin{proposition} 
\label{PropGHyperg1}
Let $ \bm{\zeta}=(\zeta _1, \ldots, \zeta_N)$ be a vector of  the zeros of the 
generalized hypergeometric polynomial $P_{N}\left( \alpha _{1},\ldots,\alpha
_{p};\beta _{1},\ldots,\beta _{q};z\right) $ defined by~\eqref{GHypergPol}. If these zeros are all different among themselves, then they
satisfy the following system of $N$ algebraic equations~\cite{BihunCalogero14-1}: 
\begin{equation}
\sum_{k=1}^{q+1}\left[ b_{k}~\ f_{n}^{\left( k\right) }\left(  \bm{%
\zeta }\right) \right] -\sum_{j=0}^{p}\left[ a_{j}~g_{n}^{\left( j\right)
}\left(  \bm{\zeta }\right) \right] =0, \;\;\;n=1,\ldots,N,  \label{Eqzitan}
\end{equation}%
where the coefficients $b_k$ and $a_j$ are defined by~\eqref{aj},~\eqref{bk}, while the functions $f_{n}^{\left( j\right)
}\left(  \bm{\zeta }\right) $ and $g_{n}^{\left( j\right)}\left(  \bm{\zeta }\right) $ are defined by~\eqref{fnj},~\eqref{gnj}.
\end{proposition}

Note that in the last Proposition, the functions $f_{n}^{\left( j\right) }\left( 
 \bm{\zeta }\right) $ and $g_{n}^{\left( j\right) }\left(  \bm{%
\zeta }\right) $ are universal in the sense that they  do not depend on the
generalized hypergeometric polynomial under consideration.

Let us now introduce the following functions of the variable $\bm{\zeta}=(\zeta_1, \ldots, \zeta_N)$:
\begin{equation}
f_{n,m}^{\left( j \right) }\left(  \bm{\zeta }\right) = \frac{%
\partial ~f_{n}^{\left( j\right) }  }{\partial \zeta_{m}}\left(  \bm{\zeta}\right), 
\;\;\;
g_{n,m}^{\left(
j\right) }\left(  \bm{\zeta }\right) = \frac{\partial
~g_{n}^{\left( j\right) }}{\partial ~\zeta_{m}}\left(  \bm{\zeta}\right) .  
 \label{fgjnm}
\end{equation}
Explicit expressions for $f_{n,m}^{\left( j \right) }\left(  \bm{\zeta }\right)$ and $g_{n,m}^{\left( j \right) }\left(  \bm{\zeta }\right)$ for small values of $j$ are reported in~\cite{BihunCalogero14-1}.

\begin{proposition} 
\label{PropGHyperg2}
Let $ \bm{\zeta}=(\zeta _1, \ldots, \zeta_N)$ be a vector of  the zeros of the 
generalized hypergeometric polynomial $P_{N}\left( \alpha _{1},\ldots,\alpha
_{p};\beta _{1},\ldots,\beta _{q};z\right) $ defined by~\eqref{GHypergPol}. If these zeros are all different among themselves, then  the $N\times N$ matrix $%
{L}\left(  \bm{\zeta }\right) $  defined componentwise by~\cite{BihunCalogero14-1}
 \begin{equation}
L_{nm}\equiv L_{nm}\left(\alpha_1, \ldots, \alpha_p; \beta_1, \ldots, \beta_q;N;  \bm{\zeta }\right) =\sum_{k=1}^{q+1}\left[
b_{k}~f_{n,m}^{\left( k\right) }\left(  \bm{\zeta }\right) \right]
-\sum_{j=1}^{p}\left[ a_{j}~g_{n,m}^{\left( j\right) }\left(  \bm{%
\zeta }\right) \right] ~,  \label{Mnmtilde}
\end{equation}%
where the coefficients $a_{j}$ and $b_{k}$ are defined by (\ref{aj}) and (\ref{bk}), while the functions $f_{n,m}^{\left( k\right)
}\left(  \bm{\zeta }\right) $ and $g_{n,m}^{\left( j\right)
}\left(  \bm{\zeta }\right) $ are defined by~\eqref{fgjnm},
has the $N$ eigenvalues 
\begin{equation}
\lambda _{m}\equiv \lambda_m\left( \beta _{1},\ldots,\beta _{q};N\right)
=m~\prod\limits_{k=1}^{q}\left( \beta _{k}-1+m\right), \;\;\;m=1,\ldots,N.  \label{mumtilde}
\end{equation}%
\end{proposition}

Note that in the last Theorem, the functions $f_{n,m}^{\left( j\right) }\left( 
 \bm{\zeta }\right) $ and $g_{n,m}^{\left( j\right) }\left(  \bm{%
\zeta }\right) $ are {universal}: they do not depend on the
generalized hypergeometric polynomial under consideration. 

The  $N\times N$ matrix $ {L}$ defined in the last Theorem depends on the zeros $\bm{\zeta}=(\zeta_1, \ldots, \zeta_N)$ of the generalized hypergeometric polynomial $P_{N}\left( \alpha _{1},\ldots,\alpha
_{p};\beta _{1},\ldots,\beta _{q};z\right) $ defined by~\eqref{GHypergPol} and, via these zeros, on the $p+q$ parameters $(\alpha _{1},\ldots,\alpha
_{p};\beta _{1},\ldots,\beta _{q})$. This matrix is isospectral because
 its eigenvalues $\lambda_m$ given by (\ref{mumtilde}) depend only on the 
parameters $\beta _{1}, \ldots, \beta_q$. Moreover, these eigenvalues are rational if the parameters $\beta _{1}, \ldots, \beta_q$ are rational, a diophantine property.

An immediate generalization of Proposition~\ref{PropGHyperg2} stems from the following observation. If in the generalized hypergeometric polynomial $Q_N(z)\equiv P_N(\alpha_1, \ldots, \alpha_p, \ldots, \alpha_{p+r}; \beta_1, \ldots, \beta_q, \ldots, \beta_{q+r};z)$  the last $r$ parameters  of the $\alpha$ and $\beta$ type are equal, that is, $\alpha_{p+j}=\beta_{q+j}$ for all $j=1, \ldots, r$, then the polynomial $Q_N(z)$ reduces to the generalized hypergeometric polynomial $P_N(z)\equiv P_N(\alpha_1, \ldots, \alpha_p; \beta_1, \ldots, \beta_q;z)$. In this setting, one may apply Proposition~\ref{PropGHyperg2} to the polynomial $Q_N(z)$ to obtain an isospectral matrix defined in terms of the zeros of the polynomial  $P_N(z)=Q_N(z)$ that is different from the matrix $L$ defined by~\eqref{Mnmtilde}. For example, by considering the polynomial $P_N(\alpha_1, \alpha_2; \beta_1, \beta_2; z)$ with $\alpha_2=\beta_2$, one can find an isospectral matrix different from~\eqref{12d} defined in terms of the zeros of the polynomial $P_N(\alpha_1; \beta_1; z)$, see Sect. 2.4 of~\cite{BihunCalogero14-1}.

\section{Zeros of generalized basic hypergeometric polynomials}

In this section, while summarizing the results of~\cite{BihunCalogero15}, we extend the method introduced in Section~\ref{PedagogicalExample} to construct an isospectral matrix defined in terms of the zeros of generalized basic hypergeometric polynomial $P_N(z) \equiv P_N(\alpha_1, \ldots, \alpha_r; \beta_1,\ldots, \beta_s;q;z)$ defined in~\eqref{GBasicHypergPol}. To this end, let us consider the differential $q$-difference equation (D$q$DE)

\begin{subequations}
\label{DqDE}
\begin{equation}
\frac{\partial }{\partial t} \psi(z,t) = \mathcal{A} \psi(z,t),
\end{equation}
where $\mathcal{A}$ is the linear $q$-difference operator
\begin{equation}
\mathcal{A}=z^{-1}\left[\Delta_1 \prod_{k=1}^s (\Delta_{\beta_k/q} )-z~\Delta_{q^{-N}} \prod_{j=1}^r (\Delta_{\alpha_j}) (\delta_q)^{s-r}\right]
\label{qDiffOpr}
\end{equation}
\end{subequations}
and note that the generalized basic hypergeometric polynomial $P_N(z)$ is a $t$-independent solution of the last D$q$DE, see \eqref{qDifferenceODEGenBasic}. The operator $\mathcal{A}$ equals the operator in the left-hand side of the $q$-difference equation~\eqref{qDifferenceODEGenBasic}, times a factor of $z^{-1}$. The presence of this factor ensures existence of monic polynomial solutions $\psi_N(z,t)$, given by~\eqref{eq1}, of  D$q$DE~\eqref{DqDE}. More precisely, because $\frac{\partial }{\partial t} \psi_N(z,t)$ is a polynomial (in $z$) of degree at most $N-1$,  for D$q$DE~\eqref{DqDE} to have a polynomial solution $\psi_N(z,t)$, $\mathcal{A} \psi_N(z,t)$  must be a polynomial of degree at most $N-1$. Let us confirm that this is indeed the case.

Recall that the operator $\Delta_\gamma$ does not raise degrees of polynomials when acting on them, see the remark following display~\eqref{DeltaOperator}. Also, the operator $\Delta_1$ annihilates functions independent of $z$, while the operator $\Delta_{q^{-N}}$ annihilates $z^N$. Therefore, the expression $$z\mathcal{A} \psi_N(z,t)=\left[\Delta_1 \prod_{k=1}^s (\Delta_{\beta_k/q} )-z~\Delta_{q^{-N}} \prod_{j=1}^r (\Delta_{\alpha_j}) (\delta_q)^{s-r}\right] \psi_N(z,t)$$ is a polynomial in $z$ of degree at most $N$ with zero constant term, rendering $\mathcal{A} \psi_N(z,t)$ to be a polynomial in $z$ of degree at most $N-1$. Upon substitution of the anzatz $\psi_N(z,t)=z^N+\sum_{n=1}^N c_n(t) z^{N-n}$ into equation~\eqref{DqDE}, we find that $\psi_N(z,t)$ solves the equation if and only if the coefficients $\bm{c}(t)=\big(c_1(t), \ldots, c_N(t)\big)$ satisfy the following linear system of ODEs:
\begin{eqnarray}
&&\dot{c}_{m}\left( t\right) =  
-\left[ q^{\left( s-r\right) \left( N-m\right) }\left( q^{-m}-1\right)
~\prod\limits_{j=1}^{r}\left( \alpha _{j}~q^{N-m}-1\right) \right]
~c_{m}\left( t\right),  \notag \\
&&+\left[ \left( q^{N-m+1}-1\right)
~\prod\limits_{k=1}^{s}\left( \beta _{k}~q^{N-m}-1\right) \right]
~c_{m-1}\left( t\right)\notag\\
&&m=1,2,\ldots,N,  \;\;\; c_0=1. \label{cSystemBasic}
\end{eqnarray}
In matrix form, the last system reads
\begin{equation}
\dot{\bm{c}}(t)=A \bm{c}(t)+\bm{h},
\label{cSystemBasicMatrixForm}
\end{equation}
where $A$ is a lower diagonal $N \times N$ matrix with the $N$ eigenvalues 
\begin{equation}
\lambda _{m}=-q^{\left( s-r\right) \left( N-m\right) }\left( q^{-m}-1\right)
~\prod\limits_{j=1}^{r}\left( \alpha _{j}~q^{N-m}-1\right), \;\;\;m=1,2,\ldots,N  \label{eigenvaluesmunBasic}
\end{equation}
and $\bm{h}$ is the $N$-vector $\bm{h}=\big((q^N-1) \prod_{k=1}^s(\beta_k q^{N-1}-1),0, \ldots,0 \big)^T$.

On the other hand, by a substitution of $\psi_N(z,t)=\prod_{n=1}^N[z-z_n(t)]$ into~\eqref{DqDE}, we obtain the following nonlinear system of ODEs for the zeros $\bm{z}(t)$:
\begin{eqnarray}
&&\dot{z}_{n}=(-1)^{s+1}\Bigg\{(q-1)f_{n}(1,\bm{z})+%
\sum_{k=1}^{s}b_{k}\frac{(-1)^{k}}{q^{k}}\Big[%
(q^{k+1}-1)f_{n}(k+1,\bm{z})-(q^{k}-1)f_{n}(k,\bm{z})\Big]%
\Bigg\}  \notag \\
&&+(-1)^{r}z_{n}\Bigg\{q^{-N}(q^{s-r+1}-1)f_{n}(s-r+1,\bm{z}%
)-(q^{s-r}-1)f_{n}(s-r,\bm{z})  \notag \\
&&+\sum_{j=1}^{r}a_{j}(-1)^{j}\Big[%
q^{-N}(q^{j+s+1-r}-1)f_{n}(j+s+1-r,\bm{z})\notag\\
&&-(q^{j+s-r}-1)f_{n}(j+s-r,%
\bm{z})\Big]\Bigg\},  \label{zSystemBasic}
\end{eqnarray}%
where 
\begin{equation}
f_{n}(p,\bm{z})=f_{n}(p,z_{1},\ldots ,z_{N})=\prod_{\ell =1,\ell \neq
n}^{N}\left( \frac{q^{p}~z_{n}-z_{\ell }}{z_{n}-z_{\ell }}\right)
,\;n=1,2,\ldots,N  \label{fnpz}
\end{equation}
and the complex coefficients $\{a_j\}_{j=1}^r$ and $\{b_k\}_{k=1}^s$ are defined as the unique set of complex numbers that satisfy the following polynomial identities:
\begin{equation}
\prod\limits_{j=1}^{r}\left( 1+\alpha _{j}~x\right) = 1+\sum_{j=1}^{r} 
a_{j} ~x^{j},
\label{ajBasic}
\end{equation}
\begin{equation}
\prod\limits_{k=1}^{s}\left( 1+\beta _{k}~x\right) = 1+\sum_{k=1}^{s} 
 b_{k} ~x^{k}.
 \label{bkBasic}
\end{equation}
We thus proved the following Lemma.

\begin{lemma} A time-dependent monic polynomial~\eqref{eq1} with distinct zeros $z_m(t)$ satisfies D$q$DE~\eqref{DqDE} if and only if its coefficients $\bm{c}(t)$ satisfy  linear system~\eqref{cSystemBasic} if and only if its zeros $\bm{z}(t)=\big(z_1(t), \ldots, z_N(t)\big)$ satisfy  nonlinear system~\eqref{zSystemBasic}. In particular, system~\eqref{zSystemBasic} is solvable in terms of algebraic operations.
\end{lemma}

Because the generalized basic hypergeometric polynomial $P_N(z)$ defined by~\eqref{GBasicHypergPol} is a $t$-independent solution of D$q$DE~\eqref{DqDE}, every vector $\bm{\zeta}=(\zeta_1, \ldots, \zeta_N)$  of the zeros  of $P_N(z)$ is an equilibrium of system~\eqref{zSystemBasic}, provided that the zeros are distinct. We thus obtain the following algebraic identities satisfied by the zeros of $P_N(z)$.

\begin{proposition} 
\label{PropGBasicHyperg1}
Let  $\bm{\zeta}=(\zeta_1, \ldots, \zeta_N)$ be a vector of zeros of the generalized basic
hypergeometric  polynomial $P_{N}\left(\alpha_1, \ldots, \alpha_r;
\beta_1, \ldots, \beta_s;q;z\right)$ defined by~\eqref{GBasicHypergPol}.  If these zeros are all different among themselves, they satisfy the following set of $N$ algebraic equations~\cite{BihunCalogero15}: 
\begin{eqnarray}
-\prod\limits_{m=1}^{N}\left( \zeta _{n}~q-\zeta _{m}\right)
+\sum_{k=1}^{s}\left( -q\right) ^{-k}b_{k}~
\left[ \prod\limits_{m=1}^{N}\left( \zeta _{n}~q^{k}-\zeta _{m}\right)
-\prod\limits_{m=1}^{N}\left( \zeta _{n}~q^{k+1}-\zeta _{m}\right) \right] &&
\notag \\
-\left( -1\right) ^{r-s}~\zeta _{n}~\left[ \prod\limits_{m=1}^{N}\left(
\zeta _{n}~q^{s-r}-\zeta _{m}\right) -q^{-N}\prod\limits_{m=1}^{N}\left(
\zeta _{n}~q^{s-r+1}-\zeta _{m}\right) \right] &&  \notag \\
-\left( -1\right) ^{r-s}~\zeta _{n}~\left\{ \sum_{j=1,~j\neq r-s}^{r}\left[
\left( -1\right) ^{j}a_{j}
~\prod\limits_{m=1}^{N}\left( \zeta _{n}~q^{s-r+j}-\zeta _{m}\right) \right]
\right. &&  \notag \\
\left. -q^{-N}\sum_{j=1,~j\neq r-s-1}^{r}\left[ \left( -1\right)
^{j}a_{j}~\prod\limits_{m=1}^{N}\left(
\zeta _{n}~q^{s-r+j+1}-\zeta _{m}\right) \right] \right\} =0, &&  \notag \\
n=1,2,..,N, &&  \label{Prop1}
\end{eqnarray}
where the sets of coefficients $\{a_{j}\}_{j=1}^r$ and $\{b_{k}\}_{k=1}^s$
are defined by identities~\eqref{ajBasic} and~\eqref{bkBasic} in terms of the parameters $\alpha_1, \ldots, \alpha_r$ and $\beta_1, \ldots, \beta_s$, respectively. 
\end{proposition}

The linearization of system~\eqref{zSystemBasic} about its equilibrium $\bm{\zeta}=(\zeta_1, \ldots, \zeta_N)$, a vector of zeros of the   polynomial ~\eqref{GBasicHypergPol}, yields the following system of  ODEs:
\begin{equation}
\dot{\bm{x}}(t)=L \bm{x}(t),
\end{equation}
where the $N\times N$ matrix $L\equiv L(\bm{\zeta})$ is given by~\eqref{LBasic}. Using an argument similar to the one employed in the proof of Theorem~\ref{Thm:IsospMatrixP11}, one may conclude that the set of eigenvalues of $L$ coincides with the set of eigenvalues of the matrix of coefficients $A$ in the linear system~\eqref{cSystemBasicMatrixForm}. Hence the following Proposition holds.

\begin{proposition} 
\label{PropGBasicHyperg2}
Let $ \bm{\zeta}=(\zeta _1, \ldots, \zeta_N)$ be a vector of  the zeros of the 
generalized basic hypergeometric polynomial $P_{N}\left( \alpha _{1},\ldots,\alpha
_{r};\beta _{1},\ldots,\beta _{s};q;z\right) $ defined by~\eqref{GBasicHypergPol}. If these zeros are all different among themselves, then  the $N\times N$ matrix $%
{L}\left(  \bm{\zeta }\right) $  defined componentwise by~\cite{BihunCalogero15}
\begin{subequations}
\label{LBasic}
\begin{eqnarray}
&&L_{nn}\equiv L_{nn}\left( \alpha _{1},\ldots,\alpha
_{r};\beta _{1},\ldots,\beta _{s};q;N;\bm{\zeta}\right)\notag\\
&&=(-1)^{s}\Bigg\{(q-1)^{2}g_{n}(1,\bm{\zeta }%
)+\sum_{k=1}^{s}b_{k}\frac{(-1)^{k}}{q^{k}}\big[%
(q^{k+1}-1)^{2}g_{n}(k+1,\bm{\zeta })-(q^{k}-1)^{2}g_{n}(k,\bm{%
\zeta })\big]\Bigg\}  \notag \\
&&+(-1)^{r+1}\zeta _{n}\Bigg\{q^{-N}(q^{s-r+1}-1)^{2}g_{n}(s-r+1,\bm{%
\zeta })-(q^{s-r}-1)^{2}g_{n}(s-r,\bm{\zeta })  \notag \\
&&+\sum_{j=1}^{r}a_{j}(-1)^{j}\big[%
q^{-N}(q^{j+s+1-r}-1)^{2}g_{n}(j+s+1-r,\bm{\zeta }%
)-(q^{j+s-r}-1)^{2}g_{n}(j+s-r,\bm{\zeta })\big]\Bigg\}  \notag \\
&&+(-1)^{r}\Bigg\{q^{-N}(q^{s-r+1}-1)f_{n}(s-r+1,\bm{\zeta }%
)-(q^{s-r}-1)f_{n}(s-r,\bm{\zeta })  \notag \\
&&+\sum_{j=1}^{r}a_{j}(-1)^{j}\Big[%
q^{-N}(q^{j+s+1-r}-1)f_{n}(j+s+1-r,\bm{\zeta }%
)-(q^{j+s-r}-1)f_{n}(j+s-r,\bm{\zeta })\Big]\Bigg\}~,  \notag \\
&&n=1,2,\ldots,N,  \label{LnnBasic}
\end{eqnarray}%
\begin{eqnarray}
&&L_{nm}\equiv L_{nm}\left( \alpha _{1},\ldots,\alpha
_{r};\beta _{1},\ldots,\beta _{s};q;N;\bm{\zeta}\right)\notag\\
&&=(-1)^{s+1}\frac{\zeta _{n}}{(\zeta _{n}-\zeta _{m})^{2}}\Bigg\{%
(q-1)^{2}f_{nm}(1,\bm{\zeta })  \notag \\
&&+\sum_{k=1}^{s}b_{k}\frac{(-1)^{k}}{q^{k}}\big[%
(q^{k+1}-1)^{2}f_{nm}(k+1,\bm{\zeta })-(q^{k}-1)^{2}f_{nm}(k,%
\bm{\zeta })\big]\Bigg\}  \notag \\
&&+(-1)^{r}\frac{\zeta _{n}^{2}}{(\zeta _{n}-\zeta _{m})^{2}}\Bigg\{%
q^{-N}(q^{s-r+1}-1)^{2}f_{nm}(s-r+1,\bm{\zeta }%
)-(q^{s-r}-1)^{2}f_{nm}(s-r,\bm{\zeta })  \notag \\
&&+\sum_{j=1}^{r}a_{j}(-1)^{j}\big[%
q^{-N}(q^{j+s+1-r}-1)^{2}f_{nm}(j+s+1-r,\bm{\zeta }%
)-(q^{j+s-r}-1)^{2}f_{nm}(j+s-r,\bm{\zeta })\big]\Bigg\},  \notag \\
&&n,m=1,2,\ldots,N,~~~n\neq m~,  \label{LnmBasic}
\end{eqnarray}
\end{subequations}
where
\begin{equation}
f_{nm}(p,\bm{\zeta})=\prod_{\ell =1,\ell
\neq n,m}^{N}\left( \frac{q^{p}~ \zeta_{n}-\zeta_{\ell }}{\zeta_{n}-\zeta_{\ell }}\right),
\label{fnmpz}
\end{equation}
\begin{equation}
g_{n}(p,\bm{\zeta})=\sum_{k=1,k\neq n}^{N}%
\left[ f_{nk}(p,\bm{\zeta})~\frac{\zeta_{k}}{(\zeta_{n}-\zeta_{k})^{2}}\right],
\label{gnpz}
\end{equation}
and the coefficients $a_{j}$ and $b_{k}$ are defined by (\ref{ajBasic}) and (\ref{bkBasic}),
has the $N$ eigenvalues 
\begin{equation}
\lambda _{n}\equiv \lambda_n\left( \alpha _{1},\ldots,\alpha
_{r};q;N\right)
=-q^{\left( s-r\right) \left( N-n\right) }\left( q^{-n}-1\right)
~\prod\limits_{j=1}^{r}\left( \alpha _{j}~q^{N-n}-1\right), \;\;\;n=1,2,\ldots,N.  \label{mun}
\end{equation}
\end{proposition}
Note that the eigenvalues $\lambda_n$ of the last matrix $L$ do not depend on the parameters $(\beta_1, \ldots, \beta_s)$ and are moreover rational if the parameters $( \alpha _{1},\ldots,\alpha_{r};q)$ are rational, a diophantine property.

\section{Zeros of Wilson and Racah polynomials}

Wilson and Racah polynomials are at the top of the Askey scheme of orthogonal polynomials~\cite{KoekoekSwarttouw98}. They are defined in terms of the generalized hypergeometric function~\eqref{GHypergF}, yet are not particular cases of the generalized hypergeometric polynomial~\eqref{GHypergPol}. In this section we summarize the results of~\cite{BihunCalogero14-2} and provide isospectral matrices defined in terms of the zeros of these polynomials constructed in.

\subsection{Wilson Polynomials}

The $N$-th degree Wilson polynomial $W_{N}(z;a,b,c,d)$ with $z=x^{2}$ is defined
by  
\begin{subequations}
\label{WilsonPol}
\begin{eqnarray}
&& W_N(z)\equiv W_{N}(z;a,b,c,d)=\left( a+b\right) _{N}~\left( a+c\right) _{N}~\left(
a+d\right) _{N} \notag\\
&& \cdot {}_{4}F_{3}\left( \left. 
\begin{array}{c}
-N,~N+a+b+c+d-1,~a+\mathbf{i}\,x,~a-\mathbf{i}\,x \\ 
a+b,~a+c,~a+d%
\end{array}%
\right\vert 1\right),  
\end{eqnarray}%
see~\cite{KoekoekSwarttouw98}, or equivalently, by
\begin{equation}
{W}_{N}(z;a,b,c,d)=\left( a+b\right) _{N}~\left( a+c\right) _{N}~\left(
a+d\right) _{N} \sum_{k=0}^{N}\left[ \frac{\left( -N\right)
_{k}~\left( N+a+b+c+d-1\right) _{k}~\left[ a;z\right] _{k}}{k!~\left(
a+b\right) _{k}\left( a+c\right) _{k}\left( a+d\right) _{k}}\right],
\label{WilsonPolSum}
\end{equation}%
see~\cite{BihunCalogero14-2}, where $a,b,c,d$ are complex parameters such that, after appropriate cancellations, the denominators in~\eqref{WilsonPolSum} do not vanish, $\mathbf{i}$ is the imaginary unit and the modified Pochhammer symbol 
\end{subequations}
\begin{eqnarray}
&&\left[ a;x^{2}\right] _{k}=\left( a+\mathbf{i}x\right) _{k}\,\left( a-%
\mathbf{i}x\right) _{k},  \notag \\
&&\left[ a;z\right] _{0}=1,  \notag \\
&&\left[ a;z\right] _{k}=\left( a^{2}+z\right) \, \left[ \left( a+1\right)
^{2}+z\right] \cdots \left[ \left( a+k-1\right) ^{2}+z\right]
,\;\;\;k=1,2,3,\ldots \notag \\
&&  \label{ModPoch}
\end{eqnarray}
Because $\left[ a;z\right] _{k}$ is a polynomial
of degree $k$ in $z$,  the Wilson
polynomial $W_{N}(z;a,b,c,d)$\textit{\ } is indeed a polynomial of degree $N$
in $z=x^2$, and of degree $2N$ in $x$.

The polynomial
\begin{equation}
w_{2N}\left( x\right) \equiv w_{2N}\left( x;a,b,c,d\right)
=W_{N}(x^{2};a,b,c,d)  \label{w2N}
\end{equation}%
satisfies the following difference equation~\cite{KoekoekSwarttouw98}:
\begin{eqnarray}
&&\Big[B\left( -x\right) (1-\delta_{\mathbf{i}}^{(+)})  +B\left( x\right)
(1-\delta_{\mathbf{i}}^{(-)})+N\left( N+a+b+c+d-1\right) 
\Big ] ~w_{2N}\left( x\right) =0,  \label{DEWilson}
\end{eqnarray}%
where $B\left( x\right) $ is defined by 
\begin{equation}
B\left( x\right) \equiv B\left( x;a,b,c,d\right) =\frac{\left( a+ \mathbf{i}\,
x\right) \left( b+ \mathbf{i}\,x\right) \left( c+ \mathbf{i}\,x\right) \left( d+ 
\mathbf{i}\,x\right) }{2\,\mathbf{i}\,x\left( 2\,\mathbf{i}\,x+ 1\right) },
\label{B+-}
\end{equation}
and the operator $\delta_\gamma^{(\pm)}$ is defined by 
\begin{equation}
 \delta_\gamma^{(\pm)}f(x)=f(x\pm \gamma).
 \label{delta_gamma^pmOperator}
\end{equation}

Consider the following differential difference equation (DDE):
\begin{subequations}
\begin{equation}
\frac{\partial \psi (x,t)}{\partial t}=\mathcal{A} \psi(x,t),
\end{equation}
where $\mathcal{A}$ is the following difference operator:
\begin{equation}
\mathcal{A}=\mathbf{i} \Big[B\left( -x\right) (1-\delta_{\mathbf{i}}^{(+)})  +B\left( x\right)
(1-\delta_{\mathbf{i}}^{(-)})+N\left( N+a+b+c+d-1\right) 
\Big ].
\end{equation}
\label{DDEWilson}
\end{subequations}
The last DDE has solutions that are $t$-dependent monic polynomials of degree $N$. Indeed, 
let $p_{\ell }\left( z\right) \equiv p_{\ell }\left(
z;a,b,c,d\right) $ be the monic version of the Wilson polynomial $W_\ell(z)$ of degree $\ell$ defined by~\eqref{WilsonPol} with $N=\ell$, and let
\begin{equation}
\psi _{2N}\left( x,t\right) =p_{N}\left( x^{2}\right) +\sum_{m=1}^{N}\left[
c_{m}\left( t\right) ~p_{N-m}\left( x^{2}\right) \right]= \prod\limits_{n=1}^{N}\left[ x^{2}-x_{n}^{2}\left( t\right) \right]   \label{psi2NWilson}
\end{equation}
be a symmetric  monic polynomial in $x$ of degree $2N$, expressed in terms of its coefficients $c_{m}\left( t\right)$ in the basis $\left\{ p_{N-m}\left( x^{2}\right)\right \}_{m=1}^N$ and its  zeros $\left\{\pm x_{n} \left( t\right)\right\}_{n=1}^N$.
A substitution of $\psi _{2N}\left( x,t\right)$ into~\eqref{DDEWilson} yields the following decoupled linear system of ODEs for the coefficients $c_m(t)$: 
\begin{equation}
\dot{c}_{m}\left( t\right) =\mathbf{i}\,m\,\left( 2N-m+\alpha _{1}-1\right)
\,c_{m}\left( t\right)
\label{cSystemWilson}
\end{equation}
and the following nonlinear system of ODEs for the zeros $x_n(t)$:
\begin{eqnarray}
&&2 x_n \dot{x}_{n}=-\mathbf{i}\, \Bigg\{\frac{D\left(
x_{n}\right) }{2\,\mathbf{i}\,x_{n}}\,\left[ \prod\limits_{m=1,~m\neq
n}^{N}\left( \frac{x_{n}^{2}-x_{m}^{2}-1-2\,\mathbf{i}\,x_{n}}{%
x_{n}^{2}-x_{m}^{2}}\right) \right]  \notag \\
&&+\Big[ \left( x_{s}\rightarrow (-x_{s})\right) \Big] \Bigg\} ,  \label{xSystemWilson}
\end{eqnarray}
where
\begin{subequations}
\label{A}
\begin{equation}
D\left( {x}_{n}\right) =\alpha
_{4}+\mathbf{i}\,\alpha _{3}\,{x}_{n}-\alpha _{2}\,{x}_{n}^{2}-\mathbf{i}\,
\alpha _{1}\,{x}_{n}^{3}+{x}_{n}^{4},  
\end{equation}
\begin{eqnarray}
&&\alpha _{1}=a+b+c+d, \notag\\
&&\alpha _{2} =ab+ac+ad+bc+bd+cd,\notag\\
&&\alpha _{3} =bcd+acd+abd+abc, \notag\\
&&\alpha _{4} =abcd
\end{eqnarray}
\end{subequations}
and the symbol $+\Big[ \left( {x}_{s}\rightarrow (-{x}_{s})\right) %
\Big] $ denotes the {addition} of everything that comes before it
(within the curly brackets), with the replacement of ${x}_{s}$ with $(-%
{x}_{s})$ for all $s=1,2,\ldots ,N$. Because system~\eqref{cSystemWilson} is explicitly solvable, we conclude that DDE~\eqref{DDEWilson} has monic symmetric polynomial solutions of the form~\eqref{psi2NWilson} and that the nonlinear system~\eqref{xSystemWilson} is solvable in terms of algebraic operations (indeed its solution is a vector function of zeros $x_n(t)$ of ${\psi}_{2N}(x,t)$).

Because the polynomial $w_{2N}(x)=W_N(x^2)$ defined by~\eqref{w2N} is a $t$-independent solution of DDE~\eqref{DDEWilson}, see~\eqref{DEWilson}, for every vector $\bm{\bar{z}}=(\bar{z}_1, \ldots, \bar{z}_N)=(\bar{x}_1^2, \ldots, \bar{x}_N^2)$ of the zeros of the Wilson polynomial $W_N(z)$ with distinct components satisfying $\bar{z}_n \neq 0$, the vector $\bm{\bar{x}}= (\bar{x}_1, \ldots, \bar{x}_N)$ is an equilibrium of system~\eqref{xSystemWilson}. By linearizing system~\eqref{xSystemWilson} about the equilibrium $\bm{\bar{x}}$, we obtain the following linear system of ODEs:
\begin{equation}
\dot{\bm{\xi}}(t)=\mathbf{i}\, L\,\bm{\xi}(t),
\end{equation}
where the $N\times N$ matrix $L\equiv L(\bm{\bar{x}})$ is given by~\eqref{LWilson}. Using an argument similar to the one employed in the proof of Theorem~\ref{Thm:IsospMatrixP11}, one may conclude that the set of eigenvalues of $L$ coincides with the set of the eigenvalues of the (diagonal) matrix of coefficients $A$ in the linear system~\eqref{cSystemWilson}. Hence the following Proposition holds.

\begin{proposition} 
\label{PropWilson}
Suppose that $\bm{\bar{z}}=(\bar{z}_1, \ldots, \bar{z}_N)=(\bar{x}_1^2, \ldots, \bar{x}_N^2)$  is a vector of zeros of Wilson polynomial $W_{N}\left(
z;a,b,c,d\right) \equiv W_{N}\left( x^{2};a,b,c,d\right) $ of degree $N$ in $
z=x^{2}$  defined by~\eqref{WilsonPol}. If these zeros are all different among themselves and such that $\bar{z}_n =\bar{x}_n^2\neq 0$ for all $n=1,\ldots, N$, then  the $N\times N$ matrix $L$ defined componentwise by~\cite{BihunCalogero14-2}
\begin{subequations}
\label{LWilson}
\begin{eqnarray}
&& L_{nn} \equiv L_{nn}(a,b,c,d;N; \bar{\bm{x}})=
\left( 2\bar{x}_{n}\right) ^{-2}\Bigg\{\left[ \frac{2D\left( \bar{x%
}_{n}\right) }{\mathbf{i}\bar{x}_{n}}+\mathbf{i}D^{\prime }\left( \bar{x}%
_{n}\right) \right] ~\prod\limits_{\ell =1,~\ell \neq n}^{N}\left( 1-\frac{%
1+2\mathbf{i}\bar{x}_{n}}{\bar{x}_{n}^{2}-\bar{x}_{\ell }^{2}}\right)  \notag
\\
&&+2D\left( \bar{x}_{n}\right) ~\sum_{m=1,~m\neq n}\frac{\mathbf{i}\bar{x}%
_{n}-\left( \bar{x}_{n}^{2}+\bar{x}_{m}^{2}\right) }{\left( \bar{x}_{n}^{2}-%
\bar{x}_{m}^{2}\right) ^{2}}\prod\limits_{\ell =1,~\ell \neq n,m}^{N}\left(
1-\frac{1+2\mathbf{i}\bar{x}_{n}}{\bar{x}_{n}^{2}-\bar{x}_{\ell }^{2}}\right)
\notag \\
&&~+\left[ \left( \bar{x}_{s}\rightarrow (-\bar{x}_{s})\right) \right] ~%
\Bigg\}, \;\;\;n=1,2,\ldots,N,
\end{eqnarray}%
\begin{eqnarray}
&&L_{nm} \equiv L_{nm}(a,b,c,d;N; \bar{\bm{x}})=
-\left( 2\bar{x}_{n}\right) ^{-2}\Bigg\{2D\left( \bar{x}_{n}\right)
~\frac{\mathbf{i}\bar{x}_{m}\left( 1+2\mathbf{i}\bar{x}_{n}\right) }{\left( 
\bar{x}_{n}^{2}-\bar{x}_{m}^{2}\right) ^{2}}\prod\limits_{\ell =1,~\ell
\neq n,m}^{N}\left( 1-\frac{1+2\mathbf{i}\bar{x}_{n}}{\bar{x}_{n}^{2}-\bar{x}%
_{\ell }^{2}}\right)  \notag \\
&&~+\left[ \left( \bar{x}_{s}\rightarrow (-\bar{x}_{s})\right) \right] ~%
\Bigg\}, \;\;\; m,n=1,2,\ldots,N, \;\;\;m\neq n,
\end{eqnarray}%
where $D\left( \bar{x}_{n}\right) $ is defined by~\eqref{A}
\end{subequations}
and
the symbol $+\left[ \left( \bar{x}_{s}\rightarrow (-\bar{x}_{s})\right) %
\right] $ denotes the addition of everything that comes before it
(within the curly brackets), with the replacement of $\bar{x}_{s}$ with $(-%
\bar{x}_{s})$ for all $s=1,2,\ldots ,N$, 
has the $N$ eigenvalues 
\begin{equation}
\lambda_m \equiv\lambda_m(a+b+c+d;N)=m\left( 2N-m+a+b+c+d-1\right) 
,\;\;\;m=1,2,\ldots,N.  \label{eq:eigM}
\end{equation}
\end{proposition}
Note that the last matrix elements $L_{nm}$ in \eqref{LWilson} are  functions of $\bar{z}_{s}=\bar{x}_{s}^{2}$ because all the terms in $L_{nm}$ that are odd functions of $\bar{x}_{s}$ cancel due to the
addition implied by the symbol $+\left[ \left( \bar{x}_{s}\rightarrow (-\bar{
x}_{s})\right) \right] $. The eigenvalues $\lambda_m$ of the last matrix $L$ depend only on the sum $a+b+c+d$ of the parameters $a,b,c,d$ as opposed to all the parameters $a,b,c,d$. These eigenvalues are moreover rational if this sum is rational, a diophantine property.

\subsection{Racah Polynomials}

The $N$-th degree Racah polynomial  is defined by 
\begin{subequations}
\label{RacahPol}
\begin{equation}
R_{N}(\lambda (x);\alpha ,\beta ,\gamma ,\delta )=_{4}F_{3}\left( \left. 
\begin{array}{c}
-N,~N+\alpha +\beta +1,~-x,~x+\gamma +\delta +1 \\ 
\alpha +1,~\beta +\delta +1,~\gamma +1%
\end{array}%
\right\vert 1\right) ,  \label{RacahPol1}
\end{equation}%
see~\cite{KoekoekSwarttouw98}, or, equivalently, by
\begin{equation}
R_{N}(\lambda (x);\alpha ,\beta ,\gamma ,\delta )=\sum_{n=0}^{N}\frac{%
(-N)_{n}(N+\alpha +\beta +1)_{n}[\lambda (x)]_{n}}{n!(\alpha +1)_{n}(\beta
+\delta +1)_{n}(\gamma +1)_{n}},  \label{RacahPolSum}
\end{equation}
where $\lambda (x)=x(x+\gamma +\delta
+1)$, $(a)_{n}$ is the Pochhammer symbol,
\begin{eqnarray}
[ \lambda (x)]_{n} &=&(-x)_{n}(x+\gamma +\delta +1)_{n}  \notag \\
&=&-\lambda (x)\left[ -\lambda (x)+(\gamma +\delta +1)+1\right] \left[
-\lambda (x)+2(\gamma +\delta +1)+2^{2}\right]  \notag \\
&&\cdots \left[ -\lambda (x)+(n-1)(\gamma +\delta +1)+(n-1)^{2}\right],
\label{[lambda]}
\end{eqnarray}
\end{subequations}
 and $\alpha ,\beta ,\gamma ,\delta $ are
complex parameters such that, after appropriate cancellations, the denominators in the finite sum~\eqref{RacahPolSum} do not vanish.

The standard definition of Racah polynomials imposes the restriction
$ \alpha+1=-M$ or $ \beta+\delta+1=-M$ or 
$\gamma + 1 = -M$ on the complex parameters $\alpha ,\beta ,\gamma ,\delta $, with $M$ being a nonnegative integer, together with the inequality $0 \leq N\leq M$.  However, in this study, none of the last diophantine relations is required or assumed.  This is because the only property of Racah polynomials used in the construction of the isospectral matrices provided below is that these polynomials satisfy difference equation~\eqref{DERacah}, which is valid even if the diophantine restrictions mentioned above do not hold.

Consider the monic polynomial
\begin{equation}
q_{2N}(x)=\frac{(\alpha +1)_{N}(\beta +\delta +1)_{N}(\gamma +1)_{N}%
}{(N+\alpha +\beta +1)_{N}}R_{N}(\lambda (x);\alpha ,\beta ,\gamma ,\delta
)  \label{q2NRacah}
\end{equation}%
of degree $2N$ in $x$
and the related polynomial
\begin{equation}
\tilde{q}_{2N}(y)=q_{2N}(y-\theta ), \label{eq:nnR8}
\end{equation}
where 
\begin{equation}
y=x+\theta,~~~\theta =\frac{\gamma +\delta +1}{2}~,  \label{ytheta}
\end{equation}
so that 
$\lambda (x)=x(x+2\theta )=y^{2}-\theta ^{2}.$
This last polynomial
satisfies the difference equation 
\begin{subequations}
\label{DERacah}
\begin{eqnarray}
\left[ \tilde{D}(y)(\delta_{1}^{(+)}-1) + \tilde{D}(-y) (\delta_{1}^{(-)}-1) 
 -N\,(N+\alpha +\beta +1) \right] \, \tilde{q}_{2N}(y)=0,
\label{eq:R9}
\end{eqnarray}
where 
\begin{eqnarray}
\tilde{D}(y)&=&\left[ 32\,y\,(2y+1)\right] ^{-1}(2%
y+\gamma +\delta +1)(2y+\gamma -\delta +1)   \notag
\\
&&\cdot (2y+2\alpha -\gamma -\delta +1)(2y+2\beta
-\gamma +\delta +1) \label{TildeD}
\end{eqnarray}
\end{subequations}
and and $\delta_1^{(\pm)}f(y)=f(y\pm1)$, see~\eqref{delta_gamma^pmOperator}. 

Let us now consider the DDE
\begin{subequations}
\label{DDERacah}
\begin{equation}
\frac{\partial }{\partial t}f(y,t)=\mathcal{A} f(y,t),
\label{DDERacah1}
\end{equation}
where
\begin{equation}
\mathcal{A}=
\mathbf{i}\left[ \tilde{D}(y)(\delta_{1}^{(+)}-1)+\tilde{D}(-y)(\delta_{1}^{(-)}-1)-N(N+\alpha +\beta +1)%
\right].  \label{OperatorARacah}
\end{equation}
\end{subequations}

The $t$-dependent symmetric monic polynomial in the $y$ variable
\begin{equation}
\tilde{\psi}_{2N}(y,t)=\tilde{q}_{2N}(y)+\sum_{m=1}^{N}c_{m}(t)\tilde{q}_{(2N-2m)}(y)=\prod_{m=1}^{N}\left[
y^{2}-y_{m}^{2}(t)\right]  \label{eq:oR9}
\end{equation}
solves the last DDE~\eqref{DDERacah} if and only if its coefficients $c_m(t)$ (with respect to the basis $\{\tilde{q}_{(2N-2m)}(y) \}_{m=1}^N$) satisfy the decoupled linear system of ODEs
\begin{equation}
\dot{c}_{m}(t)=\mathbf{i}\,m\left( m-2N-\alpha -\beta -1\right) c_{m}(t)
\label{cSystemRacah}
\end{equation}
if and only if the zeros $y_n(t)$ satisfy the nonlinear system of ODEs
\begin{eqnarray}
2y_{n}\dot{y}_{n}=-\mathbf{i}\Bigg\{\tilde{D}%
(y_{n})(2y_{n}+1)\prod_{\ell =1,\ell \neq n}^{N}\left( 1+\frac{1+2y_{n}}{%
y_{n}^{2}-y_{\ell }^{2}}\right) +[(y_{s}\rightarrow (-y_{s}))]\Bigg\},  \label{ySystemRacah}
\end{eqnarray}%
where, again, the symbol $+[(y_{s}\rightarrow (-y_{s}))]$ denotes the addition
of the expression preceding it within the curly brackets, with $y_{s}$
replaced by $(-y_{s})$ for all $s=1,2,\ldots ,N$. 
Because system~\eqref{cSystemRacah} is explicitly solvable, we conclude that DDE~\eqref{DDERacah} has symmetric monic  polynomial solutions of the form~\eqref{eq:oR9} and that the nonlinear system~\eqref{ySystemRacah} is solvable in terms of algebraic operations (indeed its solution is a vector function of zeros $y_n(t)$ of $\tilde{\psi}_{2N}(y,t)$).

Because $\tilde{q}_{2N}(y)=(\alpha +1)_{N}(\beta +\delta +1)_{N}(\gamma +1)_{N}%
[(N+\alpha +\beta +1)_{N}]^{-1}R_{N}(\lambda (y-\theta);\alpha ,\beta ,\gamma ,\delta
) $ is a $t$-independent solution of DDE~\eqref{DDERacah}, every vector of its zeros $\bar{\bm{y}}=(\bar{y}_1, \ldots, \bar{y}_N)$ is an equilibrium of the nonlinear system~\eqref{ySystemRacah}. By linearizing the last system about the equilibrium $\bar{\bm{y}}$, we obtain the linear system
\begin{equation}
\dot{\bm{\eta }}(t)=\mathbf{i}\,{L}~\bm{\eta }(t),
\end{equation}
where the $N\times N$ matrix $L\equiv L(\bar{\bm{y}})$ is given by~\eqref{LRacah}. Arguing as in the proof of Theorem~\ref{Thm:IsospMatrixP11}, one may conclude that the set of eigenvalues of $L$ coincides with the set of eigenvalues of the (diagonal) matrix of coefficients of system~\eqref{cSystemRacah}, thus the following Proposition holds.

\begin{proposition} 
\label{PropRacah}
 Suppose that $\bm{\bar{z}}=(\bar{z}_1, \ldots, \bar{z}_N)=(\bar{y}_1^2-\theta^2, \ldots, \bar{y}_N^2-\theta^2)$  is a vector of zeros of the Racah polynomial $R_{N}\left( z;\alpha ,\beta ,\gamma
,\delta \right) \equiv R_{N}\left( y^{2}-\theta ^{2};\alpha ,\beta ,\gamma
,\delta \right) $  of degree $N$ in $
z=y^{2}-\theta^2$  defined by~\eqref{RacahPol}, where $\theta =(\gamma +\delta +1)/2$. If these zeros are all different among themselves and such that $\bar{z}_n+\theta^2 =\bar{y}_n^2\neq 0$ for all $n=1,\ldots, N$, then  the $N\times N$ matrix $L$ defined componentwise by~\cite{BihunCalogero14-2}
\begin{subequations}
\label{LRacah}
\begin{eqnarray}
&&{L}_{nn}\equiv L_{nn}(\alpha, \beta, \gamma, \delta; N; \bar{\bm{y}})\notag\\
&&=\frac{1}{2}\Bigg\{\left[ \left( \frac{\tilde{D}(\bar{y}_{n})%
}{\bar{y}_{n}^{2}}-\frac{\tilde{D}^{\prime }(\bar{y}_{n})}{\bar{y}_{n}}%
\right) (1+2\bar{y}_{n})-2\frac{\tilde{D}(\bar{y}_{n})}{\bar{y}_{n}}\right]
\prod_{\ell =1,\ell \neq n}^{N}\left( 1+\frac{1+2\bar{y}_{n}}{\bar{y}%
_{n}^{2}-\bar{y}_{\ell }^{2}}\right)   \notag \\
&&+2\frac{\tilde{D}(\bar{y}_{n})}{\bar{y}_{n}}(1+2\bar{y}_{n})\sum_{m=1,m%
\neq n}^{N}\left[ \frac{\bar{y}_{n}^{2}+\bar{y}_{m}^{2}+\bar{y}_{n}}{(\bar{y}%
_{n}^{2}-\bar{y}_{m}^{2})^{2}}\prod_{\ell =1,\ell \neq n,m}^{N}\left( 1+%
\frac{1+2\bar{y}_{n}}{\bar{y}_{n}^{2}-\bar{y}_{\ell }^{2}}\right) \right]  
\notag \\
&&+[(\bar{y}_{s}\rightarrow (-\bar{y}_{s}))]\Bigg\},\;\;n=1,2,\ldots ,N
\label{LnnRacah}
\end{eqnarray}%
and 
\begin{eqnarray}
&&{L}_{nm} \equiv L_{nm}(\alpha, \beta, \gamma, \delta; N; \bar{\bm{y}})\notag\\
&&=
-\frac{1}{(\bar{y}_{n}^{2}-\bar{y}_{m}^{2})^{2}}\Bigg\{%
\frac{\bar{y}_{m}\tilde{D}(\bar{y}_{n})}{\bar{y}_{n}}(1+2\bar{y}%
_{n})^{2}\prod_{\ell =1,\ell \neq n,m}^{N}\left( 1+\frac{1+2\bar{y}_{n}}{%
\bar{y}_{n}^{2}-\bar{y}_{\ell }^{2}}\right)   \notag \\
&&+\Big[\big(\bar{y}_{s}\rightarrow (-\bar{y}_{s})\big)\Big]\Bigg\},\;\;n,m=1,2,\ldots
,N,~~~n\neq m,  \label{LnmRacah}
\end{eqnarray}
\end{subequations}
where $\tilde{D}(y)$ is defined by~\eqref{TildeD}
and again the symbol $+\Big[\big(\bar{y}_{s}\rightarrow (-\bar{y}_{s})\big)\Big]$ denotes
the addition of everything that comes before it (within the curly brackets),
with $\bar{y}_{s}$ replaced by $(-\bar{y}_{s})$ for all $s=1,2,\ldots ,N$,
 has the $N$ eigenvalues 
\begin{equation}
\lambda_m\equiv \lambda_m(\alpha+\beta;N)
=m(m-2N-\alpha -\beta -1),\;\;\;m=1,2,\ldots ,N.
\label{eq:eigMtilde}
\end{equation}
\end{proposition}

Note that the last matrix elements $L_{nm}$ are functions 
of $\bar{z}_{s}=\bar{y}_{s}^{2}-\theta ^{2}$ because
the
addition implied by the symbol $+\Big[ \big( \bar{y}_{s}\rightarrow (-\bar{%
y}_{s})\big) \Big] $ results in cancelation of all the terms odd in $\bar{y}%
_{s}$, $s=1,2,\ldots ,N$. The eigenvalues $\lambda_m$ of the last matrix $L$ depend only on the sum $\alpha+\beta$ of the parameters $\alpha, \beta$ and do not depend on the parameters $\gamma, \delta$. These eigenvalues are moreover rational if this sum is rational, a diophantine property.

\section{Zeros of Askey-Wilson and $q$-Racah polynomials}

Askey-Wilson and $q$-Racah polynomials are at the top of the $q$-Askey scheme of orthogonal polynomials~\cite{KoekoekSwarttouw98}. They are defined in terms of the generalized basic hypergeometric function~\eqref{GBasicHypergF}, yet are not particular cases of the generalized basic hypergeometric polynomial~\eqref{GBasicHypergPol}. In this section we provide isospectral matrices defined in terms of the zeros of these polynomials, which were constructed in~\cite{BihunCalogero16}.

\subsection{Zeros of Askey-Wilson polynomials} 

The Askey-Wilson polynomial $p_{N}(a,b,c,d;q;x)$ with $x=\cos  \theta$ 
is defined by 
\begin{subequations}
\label{AWPol}
\begin{eqnarray}
\label{AWPol1}
&&p_{N}(a,b,c,d;q;\cos \theta  )=\frac{\left(
ab,ac,ad;q\right) _{N}}{a^{N}}   \notag \\
&&\cdot _{4}\phi _{3}\left( \left. 
\begin{array}{c}
q^{-N},~abcd~q^{N-1},~a~\exp \left( \mathbf{i}\theta \right) ,~a~\exp \left(
-\mathbf{i}\theta \right)  \\ 
ab,~ac,~ad
\end{array}
\right\vert q;q\right),
\end{eqnarray}
see~\cite{KoekoekSwarttouw98}, or, equivalently by
\begin{eqnarray}
&&p_{N}(a,b,c,d;q;x)=\frac{\left( ab,ac,ad;q\right) _{N}~}{a^{N}}\cdot  
\notag \\
&&\cdot \sum_{m=0}^{N}\left[ \frac{q^{m}\left( q^{-N};q\right) _{m}~\left(
abcd~q^{N-1};q\right) _{m}}{\left( q;q\right) _{m}~\left( ab;q\right)
_{m}~\left( ac;q\right) _{m}~\left( ad;q\right) _{m}}~\left\{ a;q;x\right\}
_{m}\right],  \notag \\
&&  \label{AWPolSum}
\end{eqnarray}
see~\cite{BihunCalogero16}, where  the modified \textit{q}-Pochhammer symbol 
\begin{eqnarray}
&& \left\{ a;q;x\right\} _{0} =1~;  \notag \\
&& \left\{ a;q;x\right\} _{m} =\left( 1+a^{2}-2ax\right) ~\left(
1+q^{2}a^{2}-2aqx\right) \cdot \cdot \cdot \left(
1+a^{2}q^{2(m-1)}-2aq^{m-1}x\right),  \notag \\
&&m =1,2,3,\ldots  \label{ModqPoch}
\end{eqnarray}
\end{subequations}
and the complex parameters $a,b,c,d$ together with the base $q\neq 1$ are such that, after appropriate cancellations, the denominators in the sum~\eqref{AWPolSum} do not vanish.
Because $\left\{ a;q;x\right\} _{m}$ is a
polynomial of degree $m$ in $x$,
the Askey-Wilson polynomial $p_{N}(a,b,c,d;q;x)$ defined by~\eqref{AWPol} is indeed a
polynomial of degree $N$ in $x$. It is a $q$-analogue of the Wilson polynomial~\eqref{WilsonPol}.

Consider the related rational function 
\begin{equation}
Q_N(z) \equiv Q_{N}\left( a,b,c,d;q;z\right) =p_{N}\left( a,b,c,d;q;\frac{z^{2}+1}{2~z}
\right),
\label{AWRational}
\end{equation}
 and note the relations
\begin{equation}
p_{N}\left( a,b,c,d;q;x\right) =Q_{N}\left( a,b,c,d;q;x+\sqrt{x^{2}-1}
\right),
\label{AWPolRational}
\end{equation}
\begin{equation}
x=\cos \theta =\frac{z^{2}+1}{2~z}, \;\;\;\ z=e^{i\theta }=x+\sqrt{x^{2}-1}.
\label{xz}
\end{equation}
Due to  definition~\eqref{AWRational}, the  choice of the square root $\sqrt{x^{2}-1}$ from among the two possibilities is irrelevant
because in both cases $(z^{2}+1)/(2 z)=x$.

The rational function $Q_N(z)$ satisfies the following $q$-difference equation:
\begin{eqnarray}
&&\Big[
\left( q^{-N}-1\right) 
\left(1-abcd~q^{N-1}\right) 
+D\left( z\right) (1-\delta _{q})+D\left( z^{-1}\right) (1-\delta _{q^{-1}})
\Big]~Q_{N}\left(z\right) =0,  \label{AWqDE}
\end{eqnarray}
where 
\begin{equation}
D\left( z\right) \equiv D\left( a,b,c,d;q;z\right) =\frac{\left( 1-az\right)
~\left( 1-bz\right) ~\left( 1-cz\right) ~\left( 1-dz\right) }{\left(
1-z^{2}\right) ~\left( 1-qz^{2}\right) }, \label{AWD}
\end{equation}
 $\delta_qf(z)=f(qz)$ and $\delta_{q^{-1}}f(z)=f(q^{-1}z)$, see~\eqref{delta_qOperator}.

Consider the differential $q$-difference equation (D$q$DE)
\begin{subequations}
\label{DqDEAW}
\begin{equation}
\frac{\partial \Psi\left( z,t\right) }{\partial t}=\mathcal{A}\,\Psi \left(
z,t\right),  
\end{equation}
where the $q$-difference operator $\mathcal{A}$ is defined by
\begin{equation}
\mathcal{A}=\left( q^{-N}-1\right) 
\left(1-abcd~q^{N-1}\right) 
+D\left( z\right) (1-\delta _{q})+D\left( z^{-1}\right) (1-\delta _{q^{-1}}).
\end{equation}
\end{subequations}
It is clear that the rational function~\eqref{AWRational} is a $t$-independent solution of D$q$DE~\eqref{DqDEAW}. Moreover, this D$q$DE has a $t$-dependent rational solution of the form
\begin{eqnarray}
\Psi _{N}\left( z,t\right) =\sum_{m=0}^{N}\left[ c_{m}\left( t\right)
~Q_{N-m}\left( z\right) \right],
\label{AWRationalSolnDqDE}
\end{eqnarray}
where each $Q_{N-m}\left( z\right)$ is a rational function defined by~\eqref{AWRational} with $N$ replaced by $N-m$. Indeed, the rational function~\eqref{AWRationalSolnDqDE} solves D$q$DE~\eqref{DqDEAW} if and only if the coefficients $c_m(t)$ solve the decoupled linear system of ODEs
\begin{eqnarray}
&&\dot{c}_{m}\left( t\right) =q^{-N}~\left( 1-q^{m}\right) \left( 1-abcd~q^{2N-1-m}\right) ~c_{m}\left(
t\right) ~,  \notag \\
&& m =0,1,,2...,N,  \label{cSystemAW}
\end{eqnarray}
which is, of course, explicitly solvable. Note that $c_0(t)=c_0$ is constant.

Let us now perform the following change of variables in D$q$DE~\eqref{DqDEAW}:
\begin{equation}
\Psi \left( z,t\right) =\psi \left( \frac{z^{2}+1}{2~z},t\right),
\;\;\; \psi \left( x;t\right) =\Psi \left( x+ \sqrt{x^{2}-1},t\right) \label{ChangeVarDqDEAW}
\end{equation}
to recast it as 
\begin{subequations}
\label{DqDEAWx}
\begin{equation}
\frac{\partial \psi\left( x,t\right) }{\partial t}=\tilde{\mathcal{A}}\,\psi \left(
x,t\right),  
\end{equation}
where
\begin{equation}
\tilde{\mathcal{A}}=\left( q^{-N}-1\right) 
\left(1-abcd~q^{N-1}\right) 
-D\left( z\right) ~S^{(+)}-D\left( z^{-1}\right) ~S^{(-)}
+D\left( z\right)
+D\left( z^{-1}\right),
\end{equation}
$z=x+ \sqrt{x^{2}-1}$ and
\begin{equation}
S^{ (\sigma) }~f\left( x\right) =f\left( q^{\sigma
}x+\sigma \frac{1-q^{2}}{2qz}\right), \;\;\;\sigma =\pm 1.
\end{equation}
\end{subequations}
It is then clear that the polynomial 
\begin{eqnarray}
&&\psi_N(x,t)=\Psi _{N}\left( x+ \sqrt{x^{2}-1},t\right) =
\sum_{m=0}^{N}\left[ c_{m}\left( t\right)\,p_{N-m}\left( x\right) \right] \notag\\
&&=c_0\, p_N(x)+\sum_{m=1}^{N}\left[ c_{m}\left( t\right)
\,p_{N-m}\left( x\right) \right],
\label{psi_NxAW}
\end{eqnarray}
where $p_{N-m}(x)$ are Askey-Wilson polynomials, see~\eqref{AWPol},~\eqref{AWRationalSolnDqDE} and~\eqref{AWPolRational}, solves D$q$DE~\eqref{DqDEAWx}.

Let us now represent the last polynomial $\psi_N(x,t)$ in terms of its zeros $x_n(t)$:

\begin{eqnarray}
\psi_N(x,t)= C_N \prod_{n=1}^N\left[ x-x_n(t) \right],
\label{psi_NxAWZ}
\end{eqnarray}
where $C_N$ is the leading coefficient of $\psi_N(x,t)$, which does not depend on $t$ because $c_0$ is constant, see~\eqref{psi_NxAW} and the remark after display~\eqref{cSystemAW}. Upon substitution of~\eqref{psi_NxAWZ} into D$q$DE~\eqref{DqDEAW}, one obtains the following nonlinear system of ODEs:
\begin{subequations}
\label{xSystemAW}
\begin{eqnarray}
\dot{x}_{n} &=&\frac{(q-1)}{2q^{N}}\Bigg\{G(z_{n})\prod_{\ell =1,\ell \neq
n}^{N}K(z_{n},z_{\ell })+G\left( \frac{1}{z_{n}}\right) \prod_{\ell =1,\ell
\neq n}^{N}K\left( \frac{1}{z_{n}},\frac{1}{z_{\ell }}\right) \Bigg\}, \notag\\
&&n=1,\ldots ,N,
\end{eqnarray}%
where
\begin{equation}
G({z}_{n})=D({z}_{n})~\left( q{z}_{n}-\frac{1}{{z}_{n}}%
\right),  \label{eq:GG}
\end{equation}
\begin{equation}
K({z}_{n},{z}_{m})=\frac{({z}_{m}-q{z}_{n})~(q{z}_{n}%
{z}_{m}-1)}{({z}_{m}-{z}_{n})~({z}_{n}{z}_{m}-1)},
\label{eq:KK}
\end{equation}
\end{subequations} 
 and $z_{s}=\exp \left( \mathbf{i}\theta _{s}\right) $, $x_{s}=\cos
\theta _{s}$, see~\eqref{xz}.  Note that the right-hand side of~(\ref{xSystemAW}) is defined if $x_n\neq x_m$; it is an
even function of $\theta_s $, hence a function of $x_{s}$, $s=1,\ldots ,N$. System~\eqref{xSystemAW} is solvable because its solutions are vectors of the zeros of the polynomial~\eqref{psi_NxAW} whose coefficients $c_m(t)$ are components of (explicitly known) solutions of the linear system~\eqref{cSystemAW}. 

Because the Askey-Wilson polynomial $p_N(x)\equiv p_N(a,b,c,d;q;x)$ is a $t$-independent solution of D$q$DE~\eqref{DqDEAWx}, every vector $\bar{\bm{x}}=(\bar{x}_1, \ldots, \bar{x}_N)$ of its zeros is an equilibrium of system~\eqref{xSystemAW}, provided that these zeros are distinct. We may therefore linearize system~\eqref{xSystemAW} about its equilibrium $\bar{\bm{x}}=(\bar{x}_1, \ldots, \bar{x}_N)$ to obtain a linear system
\begin{equation}
\dot{\bm{\xi}}(t)=L \, \bm{\xi}(t),
\end{equation}
where the $N\times N$ matrix $L\equiv L(\bar{\bm{x}})$ is defined by~\eqref{LAW} and has the set of eigenvalues that coincides with the set of the eigenvalues of the (diagonal) coefficient matrix of the linear system~\eqref{cSystemAW} with $m=1,\ldots,N$. Therefore, the following Proposition holds.

\begin{proposition}  
\label{PropAW}
Suppose that $\bm{\bar{x}}=(\bar{x}_1, \ldots, \bar{x}_N)$  is a vector of zeros of the Askey-Wilson polynomial $p_{N}\left( a,b,c,d;q;x \right)$  of degree $N$ in $x$ defined by~\eqref{AWPol}. Let $\bar{z}_n$ be related to $\bar{x}_n$ via $\bar{x}_n=\cos \bar{\theta}_n =(\bar{z}_n^{2}+1)/(2\bar{z}_n),  \bar{z}_n=e^{i\bar{\theta}_n }=\bar{x}_n+\sqrt{\bar{x}_n^{2}-1}$. If the zeros $\bar{x}_n$ are all different among themselves, then  the $N\times N$ matrix $L$ defined componentwise by~\cite{BihunCalogero16}
\begin{subequations}
\label{LAW}
\begin{eqnarray}
&&L_{nn}\equiv L_{nn}\left( a,b,c,d,;q;N;\bm{\bar{x}}\right)   \notag
\\ \notag
&&=\frac{(q-1)}{2q^{N}}\Bigg\{\Bigg[\frac{2\bar{z}_{n}^{2}}{\bar{z}_{n}^{2}-1%
}G(\bar{z}_{n})\sum_{m=1,m\neq n}^{N}\Big(-\frac{q}{\bar{z}_{m}-q\bar{z}_{n}}%
+\frac{q\bar{z}_{m}}{q\bar{z}_{n}\bar{z}_{m}-1} \\
&&+\frac{1}{\bar{z}_{m}-\bar{z}_{n}}-\frac{\bar{z}_{m}}{\bar{z}_{n}\bar{z}%
_{m}-1}\Big)+\frac{2\bar{z}_{n}^{2}G^{\prime }(\bar{z}_{n})}{\bar{z}%
_{n}^{2}-1}\Bigg]\prod_{\ell =1,\ell \neq n}^{N}K(\bar{z}_{n},\bar{z}_{\ell
})  \notag \\
&&+\left[ \left( \bar{z}_{s}\rightarrow \frac{1}{\bar{z}_{s}}\right) \right] %
\Bigg\}
\end{eqnarray}%
and
\begin{eqnarray}
&&L_{nm}\equiv L_{nm}\left( a,b,c,d,;q;N;\bm{\bar{x}}\right)   \notag
\\ \notag
&=&\frac{(q-1)}{2q^{N}}\Bigg\{\frac{2\bar{z}_{m}^{2}}{\bar{z}_{m}^{2}-1}G(%
\bar{z}_{n})\Big[\frac{1}{\bar{z}_{m}-q\bar{z}_{n}}+\frac{q\bar{z}_{n}}{q%
\bar{z}_{n}\bar{z}_{m}-1} \\
&&-\frac{1}{\bar{z}_{m}-\bar{z}_{n}}-\frac{\bar{z}_{n}}{\bar{z}_{n}\bar{z}%
_{m}-1}\Big]\prod_{\ell =1,\ell \neq n}^{N}K(\bar{z}_{n},\bar{z}_{\ell })+%
\left[ \left( \bar{z}_{s}\rightarrow \frac{1}{\bar{z}_{s}}\right) \right] %
\Bigg\}
\end{eqnarray}
\end{subequations}
for $n,m=1,...,N~\ $with $n\neq m$, where  $G(\bar{z}_{n})$, $K(\bar{z}_{n},\bar{z}_{m})$ are defined by~\eqref{eq:GG},~\eqref{eq:KK} 
and the symbol $+\left[ \left( \bar{z}%
_{s}\rightarrow \frac{1}{\bar{z}_{s}}\right) \right] $ indicates addition of
everything that comes before it, within the curly brackets, with $\bar{z}_{s}
$ replaced by $\frac{1}{\bar{z}_{s}}$ for all $s=1,\ldots ,N$,
has the $N$ eigenvalues%
\begin{eqnarray}
&&\lambda _{n} \equiv \lambda_n(abcd;q;N)
=q^{-N}\left(
1-q^{n}\right) \left( 1-abcd~q^{2N-1-n}\right)  \notag \\
&&=\left( q^{-N}+abcd~q^{N-1}-q^{n-N}-abcd~q^{N-1-n}\right),  \notag \\
&& n =1,2,...,N.
\end{eqnarray}
\end{proposition}

Note that after the substitution $\bar{x}_{n}=\cos \bar{\theta}_{n}$ and $\bar{z}_{n}=\exp \left( \mathbf{i}
\bar{\theta}_{n}\right) $, the components $L_{nm}$ of the matrix $L$ in the last proposition become even functions of each $\bar{\theta}_s$, hence $L_{nm}$ are functions of $\bar{x}_{s}$, $s=1,\ldots,N$. The eigenvalues $\lambda_n$ of the last matrix $L$ depend only on the product $abcd$ and the basis $q$ as opposed to all the parameters $(a,b,c,d;q)$ and are moreover rational if $abcd$ and $q$ are rational, a diophantine property.

\subsection{Zeros of $q$-Racah polynomials} 

The $q$-Racah polynomial $R_{N}(\alpha ,\beta ,\gamma ,\delta;q;z)$
is defined by 
\begin{subequations}
\label{qRacahPol}
\begin{equation}
\label{qRacahPol1}
R_{N}(\alpha ,\beta ,\gamma ,\delta ;q;z)={}_{4}\phi _{3}\left( \left. 
\begin{array}{c}
q^{-N},~\alpha \beta q^{N+1},~q^{-x},~\gamma \delta q^{x+1} \\ 
\alpha q,~\beta \delta q,~\gamma q%
\end{array}%
\right\vert q;q\right),  
\end{equation}
where
\begin{equation}
z\equiv z\left( \gamma \delta ;q;x\right) =q^{-x}+\gamma \delta q^{x+1},
\label{zqRacah}
\end{equation}
see~\cite{KoekoekSwarttouw98}, or, equivalently by
\begin{eqnarray}
&&R_{N}(\alpha ,\beta ,\gamma ,\delta ;q;z)=  \notag \\
&=&\sum_{m=0}^{N}\left[ \frac{q^{m}\left( q^{-N};q\right) _{m}\left( \alpha
\beta ~q^{N+1};q\right) _{m}\left( q^{-x};q\right) _{m}\left( \gamma \delta
q^{x+1};q\right) _{m}}{\left( q;q\right) _{m}\left( \alpha q;q\right)
_{m}\left( \beta \delta q;q\right) _{m}\left( \gamma q;q\right) _{m}}\right],  \label{qRacahPolSum}
\end{eqnarray}
see~\cite{BihunCalogero16}, where  
\begin{equation}
\left( q^{-x};q\right) _{m}\left( \gamma \delta q^{x+1};q\right)
_{m}=\prod\limits_{s=0}^{m-1}\left( 1-zq^{s}+\gamma \delta q^{2s+1}\right) ~.
\end{equation}
\end{subequations}
and the complex parameters $\alpha,\beta,\gamma,\delta$ together with the base $q\neq 1$ are such that, after appropriate cancellations, the denominators in the sum~\eqref{qRacahPolSum} do not vanish.
Because $\left( q^{-x};q\right) _{m}\left( \gamma \delta q^{x+1};q\right)
_{m}$ is a
polynomial of degree $m$ in $z$,
the $q$-Racah polynomial $R_{N}(\alpha ,\beta ,\gamma ,\delta ;q;z)$ defined by~\eqref{qRacahPol} is indeed a
polynomial of degree $N$ in $z$. It is a $q$-analogue of the Racah polynomial~\eqref{RacahPol}.

The standard definition of $q$-Racah polynomials imposes the restriction
$\alpha q = q^{-M}$ or $\beta \delta q=q^{-M}$ or $\gamma q=q^{-N}$
 on the complex parameters $\alpha ,\beta ,\gamma ,\delta $, with $M$ being a nonnegative integer, together with the inequality $0 \leq N\leq M$.  However, in this study, none of the last diophantine relations is required or assumed.  This is because the only property of the $q$-Racah polynomials used in the construction of the isospectral matrices provided below is that these polynomials satisfy difference equation~\eqref{DEqRacah}, which is valid even if the diophantine restrictions mentioned above do not hold.

The \textit{q}-Racah polynomial $R_{N}(z)\equiv R_{N}(\alpha ,\beta ,\gamma
,\delta ;q;z)$ satisfies the following \textit{q}-difference equation: 
\begin{subequations}
\label{DEqRacah}
\begin{eqnarray}
&&B\left( z\right) ~R_{N}(z^{(+)})-\left[ B\left( z\right)
+D\left( z\right) \right] ~R_{N}(z)+D\left( z\right) ~R_{N}(z^{(-)})  \notag \\
&&=\left( q^{-N}-1\right) \left( 1-\alpha \beta q^{N+1}\right) ~R_{N}(z),
\label{DEqRacah1}
\end{eqnarray}%
where%
\begin{equation}
z^{ (\pm)  }=z\left( x\pm 1\right) =q^{\pm 1}z\pm \left( \frac{%
1-q^{2}}{2q}\right) \left[ z-\sqrt{z^{2}-4\gamma \delta q}\right]
\label{z+-}
\end{equation}%
and%
\begin{equation}
B\left( z\right) =\frac{\left[ 1-\alpha qZ\left( q;z\right) \right] ~\left[
1-\beta \delta qZ\left( q;z\right) \right] ~\left[ 1-\gamma qZ\left(
q;z\right) \right] ~\left[ 1-\gamma \delta qZ\left( q;z\right) \right] }{%
\left[ 1-\gamma \delta qZ^{2}\left( q;z\right) \right] ~\left[ 1-\gamma
\delta q^{2}Z^{2}\left( q;z\right) \right] }~,  \label{BqRacah}
\end{equation}%
\begin{equation}
D\left( z\right) =\frac{q~\left[ 1-Z\left( \gamma \delta q;z\right) \right] ~%
\left[ 1-\delta Z\left( \gamma \delta q;z\right) \right] ~\left[ \beta
-\gamma Z\left( \gamma \delta q;z\right) \right] ~\left[ \alpha -\gamma
\delta Z\left( \gamma \delta q;z\right) \right] }{\left[ 1-\gamma \delta
Z^{2}\left( \gamma \delta q;z\right) \right] ~\left[ 1-\gamma \delta
qZ^{2}\left( \gamma \delta q;z\right) \right] },  \label{DqRacah}
\end{equation}
with
\begin{equation}
Z\left( \gamma \delta q;z\right) =q^{x}=\frac{z+\sqrt{z^{2}-4\gamma \delta q}%
}{2\gamma \delta q}~.  \label{Z}
\end{equation}
\end{subequations}
Note that the determination of the square root in (\ref{z+-}) and (\ref{Z}) is irrelevant as long as it is consistent throughout.

Consider the differential $q$-difference equation (D$q$DE)
\begin{subequations}
\label{DqDEqRacah}
\begin{equation}
\frac{\partial \Psi\left( z,t\right) }{\partial t}=\mathcal{A}\,\Psi \left(
z,t\right),  
\end{equation}
where the $q$-difference operator $\mathcal{A}$ is defined by
\begin{equation}
\mathcal{A}=\left(
q^{-N}-1\right) \left( 1-\alpha \beta q^{N+1}\right) +B\left( z\right)(1-\Delta^{(+)})
+D\left( z\right)(1-\Delta^{(-)}) 
\end{equation}
with 
\begin{equation}
\Delta^{(\pm)}f(z)=f\big( z^{(\pm)}\big),
\end{equation}
\end{subequations}
see~\eqref{z+-}.
From~\eqref{DEqRacah1} it is clear that the $q$-Racah polynomial~\eqref{qRacahPol1} is a $t$-independent solution of D$q$DE~\eqref{DqDEqRacah}.
 Moreover, this D$q$DE has a $t$-dependent polynomial solution of the form
\begin{eqnarray}
\Psi _{N}\left( z,t\right) =\sum_{m=0}^{N}\left[ c_{m}\left( t\right)
~R_{N-m}\left( z\right) \right],
\label{qRacahPolSolnDqDE}
\end{eqnarray}
where each $R_{N-m}\left( z\right)\equiv R_{N-m}\left(\alpha, \beta, \gamma, \delta;q; z\right)$ is a $q$-Racah polynomial defined by~\eqref{qRacahPol} with $N$ replaced by $N-m$. Indeed, polynomial~\eqref{qRacahPolSolnDqDE} solves D$q$DE~\eqref{DqDEqRacah} if and only if the coefficients $c_m(t)$ solve the decoupled linear system of ODEs
\begin{eqnarray}
&&\dot{c}_{m}\left( t\right)  
=q^{-N}~\left( 1-q^{m}\right) \left( 1-\alpha \beta ~q^{2N-m+1}\right)
~c_{m}\left( t\right) ,  \notag \\
&&m =0,1,,2...,N,
 \label{cSystemqRacah}
\end{eqnarray}
which is, of course, explicitly solvable. Note that $c_0(t)=c_0$ is constant.

Let us now represent the same polynomial $\Psi _{N}\left( z,t\right)$ in terms of its zeros $z_n(t)$:
\begin{eqnarray}
\Psi_N(z,t)= C_N \prod_{n=1}^N\left[ z-z_n(t) \right],
\label{Psi_NqRacahz}
\end{eqnarray}
where $C_N$ the leading coefficient of $\Psi_N(z,t)$, which does not depend on $t$ because $c_0$ is constant, see~\eqref{qRacahPolSolnDqDE} and the remark after display~\eqref{cSystemqRacah}. Upon substitution of~\eqref{Psi_NqRacahz} into D$q$DE~\eqref{DqDEqRacah}, one obtains the following nonlinear system of ODEs:

\begin{eqnarray}
\label{zSystemqRacah}
&&\dot{z}_{n}=B\left( z_{n}\right) ~\left( z_{n}^{(+)}-z_{n}\right) ~\prod\limits_{\ell =1,~\ell \neq n}^{N}\left( \frac{
z_{n}^{(+)}-z_{\ell }}{z_{n}-z_{\ell }}\right)  \notag \\
&&
+D\left( z_{n}\right) ~\left( z_{n}^{(-)}-z_{n}\right)
~\prod\limits_{\ell =1,~\ell \neq n}^{N}\left( \frac{z_{n}^{(-)}-z_{\ell }}{z_{n}-z_{\ell }}\right) ~,  \notag \\
&&n=1,2,...,N~,
\end{eqnarray}%
where  $B\left( z\right) \equiv B\left( \alpha ,\beta ,\gamma
,\delta ;q;z\right) $ respectively $D\left( z\right) \equiv D\left( \alpha
,\beta ,\gamma ,\delta ;q;z\right) $ are defined by \eqref{BqRacah} respectively 
\eqref{DqRacah} with (\ref{Z}), and $z_{n}^{(\pm)}$ are defined by~\eqref{z+-}.
System~\eqref{zSystemqRacah} is solvable because its solutions are vectors of the zeros of the polynomial~\eqref{Psi_NqRacahz} whose coefficients $c_m(t)$ are components of (explicitly known) solutions of the linear system~\eqref{cSystemqRacah}. 

Because the $q$-Racah polynomial $R_N(z)\equiv R_N(\alpha, \beta, \gamma, \delta;q;z)$ is a $t$-independent solution of D$q$DE~\eqref{DqDEqRacah}, every vector $\bar{\bm{z}}=(\bar{z}_1, \ldots, \bar{z}_N)$ of its zeros is an equilibrium of system~\eqref{zSystemqRacah}, provided that these zeros are distinct. We may therefore linearize system~\eqref{zSystemqRacah} about its equilibrium $\bar{\bm{z}}=(\bar{z}_1, \ldots, \bar{z}_N)$ to obtain a linear system
\begin{equation}
\dot{\bm{\xi}}(t)=L \, \bm{\xi}(t),
\end{equation}
where the $N\times N$ matrix $L=L(\bar{\bm{z}})$ is defined by~\eqref{LqRacah} and has the set of eigenvalues that coincides with the set of the eigenvalues of the (diagonal) coefficient matrix of the linear system~\eqref{cSystemqRacah} with $m=1,\ldots,N$. Therefore, the following Proposition holds.

\begin{proposition}  
\label{PropqRacah}
Let $\bm{\bar{z}}=(\bar{z}_1, \ldots, \bar{z}_N)$  be a vector of  zeros of the $q$-Racah polynomial $R_N(\alpha, \beta, \gamma, \delta;q;z)$  of degree $N$ in $z$ defined by~\eqref{qRacahPol}. Let $\bar{z}_n^{(\pm)}$ be related to $\bar{z}_n$ via $z_n^{ (\pm)  }=q^{\pm 1}z_n\pm \left( \frac{
1-q^{2}}{2q}\right) \left[ z_n-\sqrt{z_n^{2}-4\gamma \delta q}\right]$, see~\eqref{z+-}.
 If the zeros $\bar{z}_n$ are all different among themselves, then  the $N\times N$ matrix $L$ defined componentwise by~\cite{BihunCalogero16}
 \begin{subequations}
\label{LqRacah}
 \begin{eqnarray}
&&L_{nn}\equiv L_{nn}\left( \alpha, \beta, \gamma, \delta ;q;N;\bm{\bar{z}}\right)=\Bigg\{B^{\prime }(\bar{z}_{n})(\bar{z}_{n}^{(+)}-\bar{z}_{n}) 
\notag \\
&&+B(\bar{z}_{n})\Big[ C^{(+)}(\bar{z}_{n})-1+(\bar{z}_{n}^{(+)}-\bar{z}%
_{n})\sum_{m=1,m\neq n}^{N}W^{(+)}(\bar{z}_{n},\bar{z}_{m})\Big] \Bigg\}%
\prod_{\ell =1,\ell \neq n}^{N}\frac{\bar{z}_{n}^{(+)}-\bar{z}_{\ell }}{\bar{%
z}_{n}-\bar{z}_{\ell }}  \notag \\
&&+\Bigg\{D^{\prime }(\bar{z}_{n})(\bar{z}_{n}^{(-)}-\bar{z}_{n})  \notag \\
&&+D(\bar{z}_{n})\Big[ C^{(-)}(\bar{z}_{n})-1+(\bar{z}_{n}^{(-)}-\bar{z}%
_{n})\sum_{m=1,m\neq n}^{N}W^{(-)}(\bar{z}_{n},\bar{z}_{m})\Big] \Bigg\}
\prod_{\ell =1,\ell \neq n}^{N}\frac{\bar{z}_{n}^{(-)}-\bar{z}_{\ell }}{\bar{%
z}_{n}-\bar{z}_{\ell }},  \notag \\
&&n=1,...,N~,  \label{Lnn}
\end{eqnarray}%
\begin{eqnarray}
&&L_{nm}\equiv L_{nm}\left( \alpha, \beta, \gamma, \delta ;q;N;\bm{\bar{z}}\right)=B(\bar{z}_{n})\left( \frac{\bar{z}_{n}^{(+)}-\bar{z}_{n}}{\bar{z}%
_{n}-\bar{z}_{m}}\right) ^{2}\prod_{\ell =1\,\ell \neq n,m}^{N}\frac{\bar{z}%
_{n}^{(+)}-\bar{z}_{\ell }}{\bar{z}_{n}-\bar{z}_{\ell }}  \notag \\
&&+D(\bar{z}_{n})\left( \frac{\bar{z}_{n}^{(-)}-\bar{z}_{n}}{\bar{z}_{n}-%
\bar{z}_{m}}\right) ^{2}\prod_{\ell =1\,\ell \neq n,m}^{N}\frac{\bar{z}%
_{n}^{(-)}-\bar{z}_{\ell }}{\bar{z}_{n}-\bar{z}_{\ell }}%
,~n,m=1,...,N,~n\neq m,  \label{Lnm}
\end{eqnarray}
\end{subequations}
where the functions $B\left( z\right) $ and $D\left(
z\right) $ are defined by \eqref{BqRacah},~\eqref{DqRacah},~\eqref{Z}, 
\begin{equation*}
C^{(\pm)}(\bar{z}_{n})=\frac{d\bar{z}_{n}^{(+)}}{d\bar{z}_{n}}=q^{\pm 1}\pm 
\frac{1-q^{2}}{2q}\left( 1-\frac{\bar{z}_{n}}{\sqrt{\bar{z}_{n}^{2}-4\gamma
\delta q}}\right) ,
\end{equation*}%
\begin{equation*}
W^{(\pm)}(\bar{z}_{n},\bar{z}_{m})=\frac{C^{(\pm)}(\bar{z}_{n})(\bar{z}%
_{n}-\bar{z}_{m})-\bar{z}_{n}^{(\pm)}+\bar{z}_{m}}{(\bar{z}_{n}-\bar{z}%
_{m})(\bar{z}_{n}^{(\pm)}-\bar{z}_{m})},
\end{equation*} has the $N$ eigenvalues%
\begin{eqnarray}
&&\lambda _{n} \equiv \lambda _{n}\left( \alpha \beta ;q;N\right)
=q^{-N}\left( 1-q^{m}\right) \left( 1-\alpha \beta ~q^{2N-m+1}\right) ~, 
\notag \\
&&n =1,2,...,N. 
\end{eqnarray}
\end{proposition}

The eigenvalues $\lambda_n$ of the last matrix $L$ depend only on the product $\alpha \beta$ and the basis $q$ as opposed to all the parameters $(\alpha, \beta, \gamma, \delta;q)$ and are moreover
rational if $\alpha \beta$ and $q$ are rational, a diophantine property.

\section{Discussion and Outlook}

We have reviewed some recent results on the properties of the zeros of several families of special polynomials: generalized hypergeometric, generalized basic hypergeometric, Wilson and Racah as well as  Askey-Wilson and  $q$-Racah polynomials. For each polynomial family $\{P_n(z)\}_{n=1}^\infty$ listed above, we have stated an $N \times N$ isospectral matrix $L$ defined in terms of the zeros of the polynomial $P_N(z)$. The eigenvalues $\{\lambda_m\}_{m=1}^N$ of the matrix $L$ are given by neat formulas that do not involve the zeros of the polynomial $P_N(z)$ and depend on fewer parameters than the polynomial $P_N(z)$. Moreover, these eigenvalues are rational in the case where some of the parameters of $P_N(z)$ are rational, a diophantine property. These isospectral matrices are obtained via construction of several solvable (in terms of algebraic operations) first order nonlinear systems of ODEs, see~\eqref{z_nSyst1Explicit}, \eqref{zSystemBasic}, \eqref{xSystemWilson}, \eqref{ySystemRacah}, \eqref{xSystemAW}, \eqref{zSystemqRacah}. Properties of these solvable systems pose an interesting subject of future studies.

Because Wilson and Racah polynomials are at the top of the Askey scheme, while   Askey-Wilson and  $q$-Racah polynomials are at the top of the $q$-Askey scheme, it must be possible, by taking appropriate limits, to use the results of Propositions~\ref{PropWilson}, \ref{PropRacah}, \ref{PropAW}, \ref{PropqRacah} to obtain new isospectral matrices defined in terms of the zeros of all the other polynomials in the Askey and the $q$-Askey schemes. This is another topic for future investigation.

In each among the Propositions~\ref{PropGHyperg2},~\ref{PropGBasicHyperg2},~\ref{PropWilson}, \ref{PropRacah}, \ref{PropAW}, \ref{PropqRacah}, the $N$ eigenvalues of the isospectral matrix $L$ are given, while the corresponding eigenvectors and the transition matrix $T$ that diagonalizes the matrix $L$ via the similarity transformation $T^{-1}L T$ are not known. Finding similarity matrices for each isospectral matrix $L$ in Propositions~\ref{PropGHyperg2},~\ref{PropGBasicHyperg2},~\ref{PropWilson}, \ref{PropRacah}, \ref{PropAW}, \ref{PropqRacah} would provide a tool for construction of additional algebraic identities satisfied by the zeros of the appropriate polynomials. A method that compares the spectral and the pseudospectral matrix representations of linear differential operators has been employed in~\cite{AliciTaseli15, Bihun17, BihunMourning18} to construct new and re-obtain known algebraic identities satisfied by classical, Krall and Sonin-Markov orthogonal polynomials. It would be interesting to adapt the last method to the cases of the generalized hypergeometric, generalized basic hypergeometric, Wilson and Racah as well as  Askey-Wilson and  $q$-Racah polynomials and to compare the results with those reviewed in this chapter as well as to utilize the new developments to study properties of the solvable nonlinear systems~\eqref{z_nSyst1Explicit}, \eqref{zSystemBasic}, \eqref{xSystemWilson}, \eqref{ySystemRacah}, \eqref{xSystemAW}, \eqref{zSystemqRacah}.

\section*{Acknowledgements}

Results reviewed in this chapter were obtained together with Francesco Calogero, in particular during several visits to the ``La Sapienza'' University of Rome, for which hospitality the author is grateful.


\begin{thebibliography}{99}

\bibitem{Ahmed78} Ahmed S, Novel properties of the zeros of Laguerre polynomials, {\it Lettere Nuovo Cimento} {\bf 22} (1978) 371--375.

\bibitem{Ahmed79} Ahmed S, A general technique to obtain nonlinear equations for the zeros of classical orthogonal polynomials, {\it Lettere Nuovo Cimento} {\bf 26} (1979) 285--288.

\bibitem{ABCOP1979}  Ahmed S, Bruschi M, Calogero F, Olshanetsky M A, 
Perelomov A M, Properties of the zeros of the classical polynomials and
of Bessel functions, \textit{Nuovo Cimento} \textbf{49B}  (1979) 173-199.

\bibitem{AliciTaseli15} Alici H, Ta\c{s}eli H, Unification of Stieltjes-Calogero type relations for the zeros of classical orthogonal polynomials, {\it Math. Meth. Appl. Sci.}, 
\textbf{38}(14) (2015) 3118--3129.

\bibitem{BealsSzmigielski13} Beals R, Szmigielski J, Meijer $G$-functions: a gentle introduction, {\it Notices of the AMS} {\bf 60}(7) (2013) 866-872.

\bibitem{GR1990} Gasper G, Rahman M, \textit{Basic Hypergeometric Series}, Cambridge University Press, 1990.

\bibitem{Bihun17}  Bihun O, New properties of the zeros of Krall polynomials, \textit{J. Nonlinear Math. Phys.} {\bf 24}(4) (2017) 495--515.

\bibitem{BihunCalogero14-1} Bihun O, Calogero F, Properties of the zeros of generalized hypergeometric polynomials, {\it J. Math. Analysis Appl.} \textbf{419}(2) (2014) 1076--1094.

\bibitem{BihunCalogero14-2} Bihun O, Calogero F, Properties of the zeros of the polynomials belonging to the Askey scheme,
{\it Lett. Math. Phys.} \textbf{104}(12) (2014) 571--588.

\bibitem{BihunCalogero15} Bihun O, Calogero F, Properties of the zeros of generalized basic hypergeometric polynomials, {\it J. Math. Phys.}, \textbf{56} (2015) 112701, 1--15.

\bibitem{BihunCalogero16} Bihun O, Calogero F, Properties of the zeros of the polynomials belonging to
the \textit{q}-Askey scheme, {\it J. Math. Analysis Appl.} \textbf{433}(1) (2016) 525--542.


\bibitem{BihunCalogero16-1} Bihun O, Calogero F, Novel \textit{solvable} many-body problems, \textit{J. Nonlinear Math. Phys.} {\bf 23}(2) (2016) 190--212.

\bibitem{BihunCalogero16-2} Bihun O, Calogero F, A new solvable many-body problem of goldfish type, \textit{J. Nonlinear Math. Phys.} {\bf 23}(1) (2016)  28--46.

\bibitem{BihunCalogero16-3} Bihun O, Calogero F, Generations of monic polynomials such that the coefficients of the 
polynomials of the next generation coincide with the zeros of the polynomials 
of the current generation, and new solvable many-body problems, \textit{Lett. Math. Phys.} {\bf 106}(7) (2016) 1011--1031.

\bibitem{BihunCalogero17} Bihun O, Calogero F, Generations of solvable discrete-time dynamical systems, \textit{J. Math. Phys.} {\bf 58} (2017) 052701, 21 pages. 

\bibitem{BihunMourning18} Bihun O, Mourning C, Generalized Pseudospectral Method and Zeros of Orthogonal Polynomials, \textit{Adv. Math. Phys.} Vol. 2018 (2018), Article ID 4710754, 10 pages.

\bibitem{CalogeroBruschi16} Bruschi M, Calogero F, A convenient expression of the time-derivative ,
of arbitrary order $k$, of the zero $z_n(t)$ of a time dependent
polynomial $p_N(z;t)$ of arbitrary degree $N$
in $z$, and solvable dynamical systems, {\it J. Nonlinear Math. Phys.}  {\bf 23}(4) 474--485.

\bibitem{Calogero71} Calogero F, Solution of the one-dimensional N-body problem with quadratic and/or inversely quadratic pair potentials, \textit{J.
Math. Phys.} {\bf 12} (1971) 419-436; "Erratum", ibidem 37,
3646 (1996).

\bibitem{Calogero77-1} Calogero F, On the zeros of Hermite polynomials, {\it Lettere Nuovo Cimento} {\bf 20} (1977) 489--490.

\bibitem{Calogero77-2} Calogero F, On the zeros of the classical polynomials, {\it Lettere Nuovo Cimento} {\bf 19} (1977) 505--508.

\bibitem{Calogero01} Calogero F, The ``neatest" many-body problem amenable to
exact treatments (a ``goldfish"?), \textit{Physica D}  \textbf{152-153}
(2001) 78-84.

\bibitem{Calogero2001} Calogero F, {\it Classical many-body problems amenable
to exact treatments}, Lecture Notes in Physics Monographs \textbf{m66},
Springer, Heidelberg, 2001.

\bibitem{Calogero08} Calogero F, \textit{Isochronous systems}, Oxford University
Press,  2012.

\bibitem{Calogero15} Calogero F, New solvable variants of the
goldfish many-body problem, \textit{Studies Appl. Math.} 
{\bf 137}(1) (2016) 123--139.



\bibitem{Chihara78} Chihara T, {\it An introduction to orthogonal polynomials}, Gordon and Breach, 1978.

\bibitem{DriverJordaan16} Driver K, Jordaan K, Zeros of quasi-orthogonal Jacobi polynomials, {\it SIGMA} {\bf 12} (2016) 042, 13 pages.

\bibitem{Erdelyi53} {\it Higher Transcendental Functions}, Volumes 1, 2, Bateman Project, Editor Erd\'{e}lyi A, McGraw-Hill, New York, 1953.

\bibitem{Funaro92}  Funaro D, {\it Polynomial approximations of differential equations}, Springer-Verlag, Berlin, 1992.

\bibitem{Ismail00-1}  Ismail M E H, An electrostatic model for zeros of general orthogonal polynomials,
{\it Pacific J. Math.} {\bf 193} (2000) 355--369.

\bibitem{Ismail00-2} Ismail M E H, More on electrostatic models for zeros of orthogonal polynomials, {\it Numer.
Funct. Anal. Opt.} {\bf 21} (2000) 191--204.

\bibitem{KoekoekSwarttouw98} Koekoek R, Swarttouw R F, {\it The Askey-scheme of hypergeometric orthogonal polynomials and its q-analogue}, Delft University of Technology, Faculty of Technical Mathematics and Informatics, Report no. 94-05  (1994), revised in Report no. 98-17, 1998, available online at http://homepage.tudelft.nl/11r49/askey/.

\bibitem{KoekoekLeskySwarttouw10} Koekoek R, Lesky P A, Swarttouw R F, {\it Hypergeometric orthogonal polynomials and their $q$-
analogues}, Springer, 2010.

\bibitem{MMFMG07} Marcell\'{a}n F, Mart\'{i}nez-Finkelshtein A, Mart\'{i}nez-Gonz\'{a}lez P, Electrostatic models for zeros of polynomials: Old, new, and some open problems, {\it J. Computational Applied Math} {\bf 207}(2) (2007) 258--272.

\bibitem{MastroianniMilovanovic08}
Mastroianni G, Milovanovi\'c G, {\it Interpolation processes: basic theory and applications}, Springer, 2008.

\bibitem{Moser75} Moser J, Three
integrable Hamiltonian systems connected with isospectral deformations,
\textit{Adv. Math.} {bf 16} (1975) 197-220.

\bibitem{NikiforovUvarov88} Nikiforov A F, Uvarov V B, {\it Special functions of mathematical physics}, Birkh\"{a}user, 1988.

\bibitem{Phillips03} Phillips G, {\it Interpolation and approximation by polynomials}, Canadian Mathematical Society, 2003.

\bibitem{Sasaki15} Sasaki R, Perturbations around the zeros of classical orthogonal polynomials, {\it J. Math. Phys.}, \textbf{56} (2015) 042106.

\bibitem{Stieltjes1885-1}
Stieltjes T J, Sur quelques th\'{e}or\`{e}mes d'alg\`{e}bre, {\it Comptes Rendus de l'Acad\'{e}mie des Sciences} {\bf 100} (1885) 439--440.

\bibitem{Stieltjes1885-2} Stieltjes T J, Sur les polyn\^{o}mes de Jacobi, {\it Comptes Rendus de l'Académie des Sciences} {\bf 100} (1885) 620--622.

\bibitem{Stieltjes1885-3} Stieltjes T J, Sur certains polyn\^{o}mes que v\'{e}rifient une\'{e}quation diff\'{e}rentielle lin\'{e}aire
du second ordre et sur la th\'{e}orie des fonctions de Lam\'{e}, {\it Acta Math.} {\bf 6} (1885) 321--326.

\bibitem{Szego39} Szeg\"{o} G., {\it Orthogonal polynomials}, American Mathematical Society, 1939.

\bibitem{Veselov01} Veselov A P, On Stieltjes relations, Painlev\'{e}-IV hierarchy and complex monodromy, {\it J. Physics A: Mathematical and General} {\bf 34} 2001
3511--3519.

\end{thebibliography}
\end{document}